\documentclass[prx,twocolumn,floatfix,superscriptaddress,longbibliography,nofootinbib]{revtex4-2}
\usepackage{amsfonts,mathrsfs,amsmath,amsthm,amssymb,bbold}
\usepackage{amsmath}
\usepackage{bm}
\usepackage{epsfig}
\usepackage{bm,mathrsfs}
\usepackage{lineno,hyperref}
\usepackage[margin=25mm]{geometry} 
\usepackage{array} 
\usepackage{color}
\usepackage{xcolor,colortbl} 
\usepackage{soul}
\usepackage{afterpage}
\usepackage{capt-of}
\usepackage{makecell}
\usepackage{graphicx}
\usepackage[normalem]{ulem}
\usepackage{cancel}
\usepackage{siunitx}
\usepackage[version=4]{mhchem}
\DeclareSIUnit\gauss{Gauss}
\DeclareSIUnit\muB{\ensuremath{\mu_{\mathrm{B}}}}
\DeclareSIUnit\oersted{Oe} 
\usepackage{braket}
\usepackage{array}
\setcellgapes{4pt}
\newcolumntype{P}[1]{>{\arraybackslash}p{#1}}
\newcolumntype{C}[1]{>{\centering\arraybackslash}p{#1}}

\begin{document}  
\title {\bf Anisotropic Piezomagnetism in Noncollinear Antiferromagnets} 

\author{Vu Thi Ngoc Huyen}
\thanks{These authors contributed equally to this work. Contact author: vu.thi.ngoc.huyen.b6@tohoku.ac.jp} 
\affiliation{Institute for Materials Research, Tohoku University, Sendai, Miyagi 980-8577, Japan}

\author{Yuki Yanagi}
\thanks{These authors contributed equally to this work. Contact author: vu.thi.ngoc.huyen.b6@tohoku.ac.jp} 
\affiliation{
Liberal Arts and Sciences, Toyama Prefectural University, Toyama 939-0398, Japan}

\author{Michi-To Suzuki}
\affiliation{Department of Materials Science, Graduate School of Engineering, Osaka Metropolitan University, Sakai, Osaka 599-8531, Japan}
\affiliation{Center for Spintronics Research Network, Graduate School of Engineering Science,
Osaka University, Toyonaka, Osaka 560-8531, Japan}

\date{\today}

\begin{abstract}
In 3d-electron magnetic systems, the magnetic structures that transform each other by spin rotation have very close degenerate energies due to small spin-orbit coupling and can be easily controlled by chemical substitution and external magnetic fields. 
We investigate anisotropic piezomagnetic effects, exhibiting the different magnetic responses depending on the type of strain and the magnetic structures, for non-collinear magnetic states in Mn$_3A$N ($A=$ Ni, Cu, Zn, Ga) and Mn$_3X$ ($X$= Sn and Ge) based on detailed symmetry analysis using spin group and magnetic group and first-principles calculations of piezomagnetic responses. 
In Mn$_3A$N, magnetization develops along two distinct directions under the same applied stress, corresponding to two AFM states connected by spin rotation. 
Analysis of the piezomagnetic tensor based on magnetic and spin point groups for the states with and without spin-orbit coupling, respectively, shows that the difference in the magnitude of magnetization along different directions is attributed to the spin-orbit coupling. 
Mn$_3X$ are known to stabilize different AFM structures in the directions of the applied in-plane magnetic fields. 
Under uniaxial stress along the orthorhombic $x$ and $y$ axes, magnetization is induced without breaking the magnetic symmetry, but it develops in the opposite direction due to exchange interaction.
Our study demonstrates that the direction and sign of strain-induced magnetization in Mn$_3A$N and Mn$_3X$ can be effectively controlled by strain in combination with magnetic fields. 
These findings highlight the potential for strain-tunable magnetic devices in noncollinear AFMs.
\end{abstract}
\maketitle
\section{Introduction}\label{secintro}
Exploring different forms of magnetoelectric control of magnetization is essential for developing non-volatile random access memory~\cite{zhang2024electric}. 
In recent years, a promising avenue in energy efficiency research has emerged, showcasing the potential of electric field manipulation in certain materials. 
This method, either directly or indirectly influenced by elastic stress,
has demonstrated a remarkable ability to minimize energy wastage compared to conventional spin-transfer torque mechanisms. 
Referenced studies underscore its effectiveness, positioning it as a viable contender alongside current semiconductor field-effect transistors \cite{pmtheory2015,pmtheory2021}. 
Furthermore, the exploration of magnetization control through elastic stress, known as the piezomagnetic effect, presents a compelling frontier. Particularly intriguing is its application in antiferromagnetic (AFM) systems,
which are renowned for their potential in high-density storage and energy efficiency~\cite{Baltz_RevModPhys.90.015005}. 
While piezomagnetism has been associated with collinear insulating antiferromagnets such as $\alpha$--Fe$_2$O$_3$~\cite{Philips1967,Dzialoshinskii1958} and CoF$_2$~\cite{Dzialoshinskii1958,MORIYA195973}, recent experiments have revealed significant piezomagnetic effects in metallic noncollinear AFM systems,
including Mn$_3$NiN~\cite{exp2018pmmn3nin} and Mn$_3X$ ($X$ = Sn and Ge)~\cite{exp2022pmahc,expmn3x2022}.
More recently, Meng {\it et al.} demonstrated that the distortion of the triangular lattice and field-induced spin twisting induce piezomagnetism and magnetostriction in Mn$_3$Sn \cite{Meng2024}.

The AFM structures of these compounds are characterized as higher-rank magnetic multipoles~\cite{cmp2017, cmpgeneration}. The specific responses for the AFM states have been studied both theoretically~\cite{cmp2017,ahe} and experimentally~\cite{expmn3sn2015}.
Such AFM compounds, including antiperovskite Mn$_3$GaN, Mn$_3$ZnN, Mn$_3$NiN~\cite{pmtheory2015,pmtheory2021,pahc,exp2018pmmn3nin,Samathrakis2020,PhysRevMaterials.3.024407} and Mn$_3$Sn~\cite{exp2022pmahc,expmn3sn2015,expmn3x2018,expmn3x2022,cmp2017,tomiyoshi} exhibit both metallic properties and piezomagnetism, making them promising candidates for spintronic devices.
Mn$_3$Sn has also attracted renewed interest in the context of altermagnet~\cite{Cheong2024}, a kind of antiferromagnet showing intriguing physical properties such as the momentum-dependent spin splitting of band structures~\cite{Noda_C5CP07806G,Ahn_PhysRevB.99.184432,Naka_NatCommun.10.4305,Hayami_JPSJ.88.123702,Yuan_PhysRevB.102.014422,Mazin_PhysRevX.12.040002} and the anomalous Hall effect with negligibly small net magnetization~\cite{Solovyev_PhysRevB.55.8060,Smejkal_sciadv.aaz8809,Liu_PhysRevX.12.021016,Kawamura_PhysRevLett.132.156502}. 
Piezomagnetic effect is one of the characteristic features of altermagnets~\cite{Hayami_JPSJ.88.123702,HYMa2021,Spaldin_PhysRevX.14.011019,Aoyama_PhysRevMaterials.8.L041402}. 
Importantly, the magnetic structures in these materials can be controlled by varying the chemical composition or applying external magnetic fields. Theoretical predictions reflect this controllability of magnetic structures, showing that these materials can stabilize different AFM configurations that are connected by spin rotations. 
Since these spin-rotated AFM states are separated by only small energy differences~\cite{ahe}, originating from the weak spin-orbit coupling (SOC) of Mn atoms, the magnetic responses can be readily tuned through such chemical and physical perturbations.
Interestingly, the nearly degenerate AFM states can exhibit considerably different physical properties. For example, the spin-rotated AFM states observed in Mn$_3A$N include magnetic structures that exhibit anomalous Hall effects and those that do not~\cite{ahe}, as discussed later. 

In this paper, we investigate the anisotropic piezomagnetic response in such AFM states, Mn$_{3}A$N ($A$ = Ni, Cu, Zn, Ga) and Mn$_{3}X$ ($X$ = Sn and Ge), through first-principles calculations and magnetic symmetry analysis.
In symmetry analysis, spin point group theory, which provides a proper description of the magnetic symmetry in the SOC--free systems~\cite{LITVIN1974538,Litvin:a14103}, as well as ordinary magnetic point group theory, is employed to elucidate the effects of SOC for piezomagnetic effects~\cite{Watanabe_PhysRevB.109.094438}.  
The organization of the paper is as follows. 
Section~\ref{sec:method} describes the investigation methodology, including the process of studying the piezomagnetic effect through the optimization of the stress tensor and the computational details.
Sec.~\ref{sec:results} A and B analyze the piezomagnetic effects in Mn$_3A$N and Mn$_3X$, respectively, and exemplify the anisotropic response of piezomagnetism in the noncollinear AFM states.
Finally, Sec.~\ref{conclusion} contains a summary of this work.
\section{Investigation methodology} \label {sec:method}
\subsection{Symmetry argument of piezomagnetic effect} \label {sec:formu}
\begin{figure} 
\centering 
\includegraphics[width=8cm]{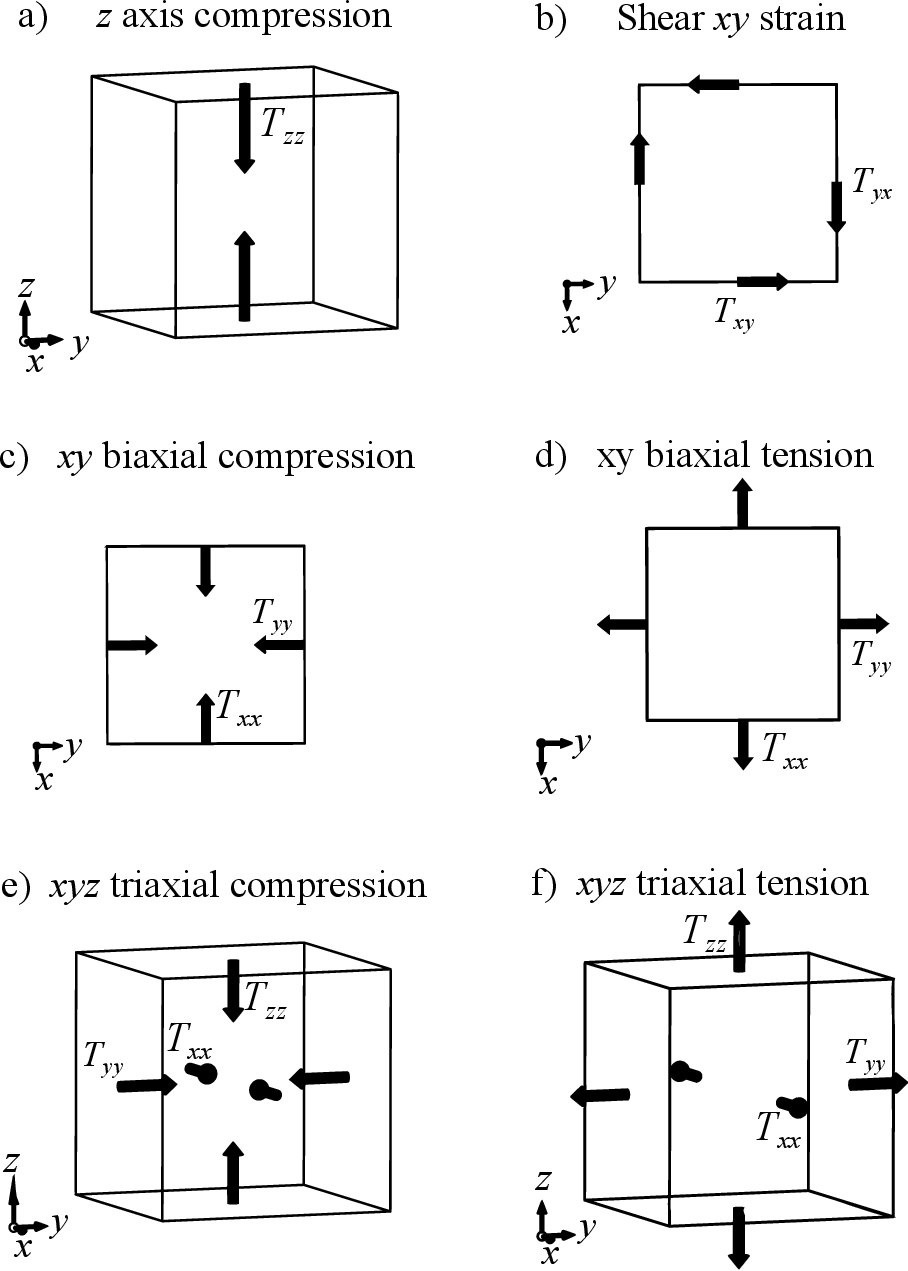}\\
\captionof{figure}{Illustration of the type of stress corresponding to each stress tensor component}
\label{fig:illustrate}
\end{figure}

To make this paper self-contained, we first briefly overview the symmetry properties of piezomagnetic effects. The piezomagnetic effect is a phenomenon in which magnetization is induced in a crystal under stress~\cite{itcD157}. It needs a third-rank tensor to relate the magnetization vector $\bm{\mathit{M}}$ and the second-rank stress tensor $\bm{\mathit{T}}$, called the piezomagnetic tensor $\bm{\mathit{Q}}$, as follows:
\begin{equation} \label{eqpiezomag}
\bm{\mathit{M=QT}},
\end{equation}
or in the index form:
\begin{equation} \label{eqpiezomagi}
M_i=Q_{ijk}T_{jk}.
\end{equation} 
Therein, $i$, $j$, and $k$ specify $x$, $y$, and $z$ in Cartesian coordinates, and $i$ indicates a component of the magnetization vector, $j$ indicates the face to which the stress is applied, and $k$ indicates the axis along which the stress is projected. The diagonal elements of the stress tensor $T_{ii}$ represent the normal stress components for the $i$--plane, while the off-diagonal elements $T_{ij} (i\neq j)$ represent the shear stress components. 
The stress tensor is symmetric in Cartesian coordinates~\cite{itcD132,itcD157}, meaning that $T_{ij}=T_{ji}$. This symmetry allows the stress tensor to be reduced from its nine-component form to a six-component form. Table~\ref{tab:9t6} shows the relationship between the six-component and nine-component forms of the stress and piezomagnetic tensor components.
\begin{table}
\captionof{table}{Conversion of indexes of stress and piezomagnetic tensors}
\label{tab:9t6}
(a) Stress tensor components \\
\begin{tabular}{P{1.5cm}P{0.6cm}P{0.6cm}P{0.6cm}P{1.3cm}P{1.3cm}P{1.2cm}} 
\hline
\hline
9 indexes& $T_{xx}$ & $T_{yy}$& $T_{zz}$& $T_{yz}, T_{zy}$ & $T_{xz}, T_{zx}$ & $T_{xy}, T_{yx}$  \\
 \hline
6 indexes & $T_1$ & $T_2$ & $T_3$ & $T_4$ & $T_5$ &$T_6$ \\
\hline
\hline
\end{tabular}  
(b) Piezomagnetic tensor components \\
\begin{tabular}{P{2.6cm}P{2.6cm}P{2.5cm}} 
\hline
\hline
 $Q_{xxx} \equiv Q_{11}$ & $Q_{yxx} \equiv Q_{21}$ & $Q_{zxx} \equiv Q_{31}$  \\
 $2Q_{xxy} \equiv Q_{16}$ & $2Q_{yxy} \equiv Q_{26}$ & $2Q_{zxy} \equiv Q_{36}$  \\
 $2Q_{xxz} \equiv Q_{15}$ & $2Q_{yxz} \equiv Q_{25}$ & $2Q_{zxz} \equiv Q_{35}$  \\
 $2Q_{xyz} \equiv Q_{14}$ & $Q_{yyy} \equiv Q_{22}$ & $Q_{zyy} \equiv Q_{32}$  \\
 $Q_{xzz} \equiv Q_{13}$ & $2Q_{yyz} \equiv Q_{24}$ & $2Q_{zyz} \equiv Q_{34}$  \\
 $Q_{xyy} \equiv Q_{12}$ & $Q_{yzz} \equiv Q_{23}$ & $Q_{zzz} \equiv Q_{33}$  \\
\hline
\hline
\end{tabular}  
\end{table}
We illustrate some types of stress in Fig.~\ref{fig:illustrate}.
Therein, shear stress is parallel to the applied surface, and other stresses are components perpendicular to the applied surface.

Considering the matrix elements of the piezomagnetic tensor, they must be invariant under all symmetry operations that keep the magnetic crystal invariant. The components of the third-rank tensor $Q_{ijk}$ transform under an operation $R$ as:
\begin{align} 
\tilde{Q}_{ijk}=D_{ii'}(R)D_{jj'}(R)D_{kk'}(R)Q_{i'j'k'},\label{transform}
\end{align}
where $D_{ij}(R)$ is the component of a transformation matrix for the operation $R$. If the system is invariant under an operation $R$, the relation $\tilde{Q}_{ijk}=Q_{ijk}$ is satisfied. Let us here discuss two of the most fundamental symmetries, i.e. the spatial inversion and time-reversal symmetries.
The piezomagnetic tensor $\bm{Q}$ is even under the spatial inversion operation ($\mathcal{P}$) since both the magnetization $\bm{M}$ and the stress tensor $\bm{T}$ are invariant for that operation.
Magnetization is naturally prohibited for the magnetic point groups which includes the time-reversal operation ($\mathcal{T}$).  Even if the magnetic point group contains no pure $\mathcal{T}$ symmetry, magnetization is still prohibited if the magnetic point group includes symmetry combining spatial inversion and time-reversal operations, $\mathcal{PT}$.
Since the $\mathcal{T}$ and $\mathcal{PT}$ symmetries cannot be broken by applying any strain, the piezomagnetic effect is prohibited in materials with these symmetries.
Therefore, the piezomagnetic effect must be absent for the materials which belong to following 21 magnetic point groups that preserve $\mathcal{PT}$ symmetry even without $\mathcal{T}$ symmetry, $\bar{1}'$, $2/m'$, $2'/m$, $m'm'm'$, $mmm'$, $4/m'$, $4'/m'$, $4/m'm'm'$, $4/m'mm$, $4'/m'm'm$, $\bar{3}$, $\bar{3}'$, $\bar{3}'m'$, $6/m'$, $6'/m$, $6/m'm'm'$, $6/m'mm$, $6'/mm'm$, $m'3$, $m'3m'$, and $m'3m$, as well as the ones preserving $\mathcal{T}$ symmetry.
The cubic magnetic point groups $432$, $\bar{4}3m$, and $m\bar{3}m$ also prohibit the piezomagnetic effect even with no anti-unitary symmetry since the unitary symmetry prohibits it.
%
\subsection{Optimization of stress tensor} \label {sec:optimize}
%
\begin{figure} 
\centering 
\includegraphics[width=6cm]{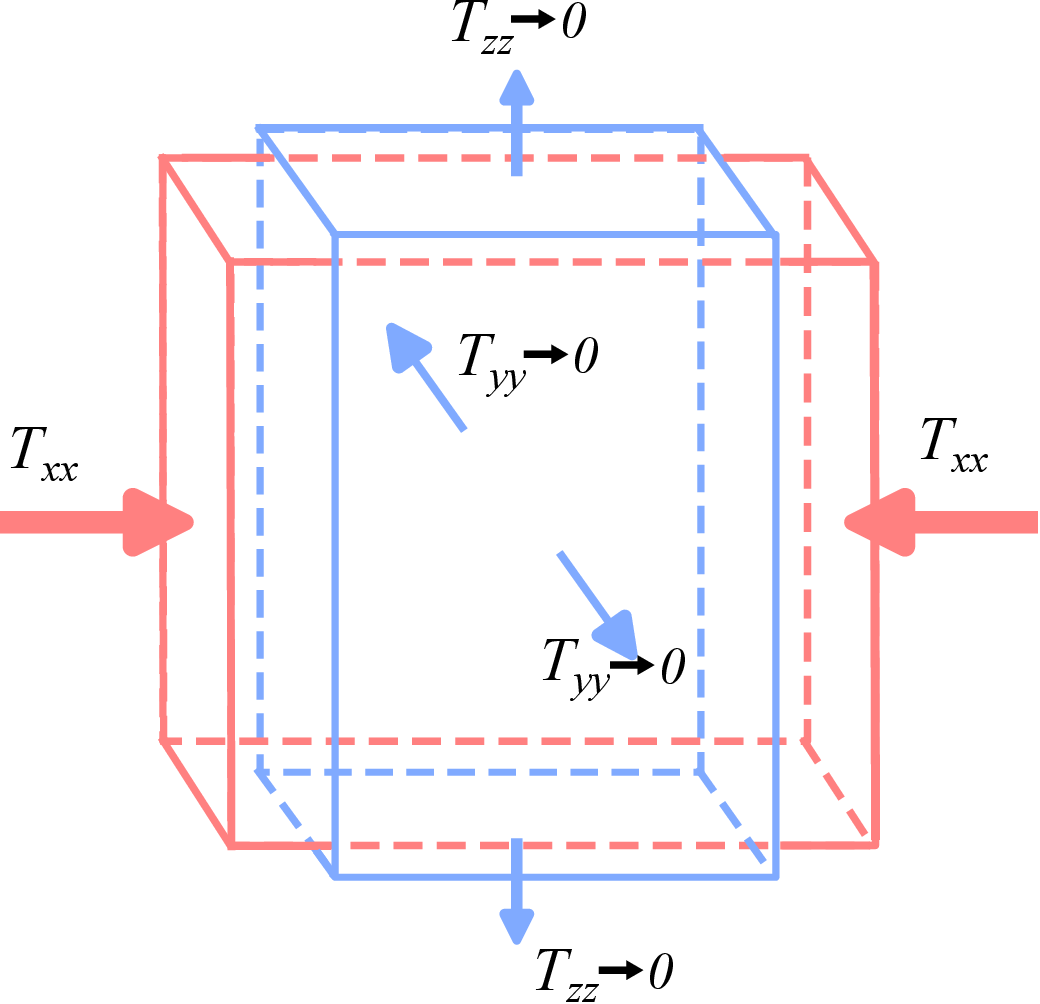}\\
\captionof{figure}{Illustration of uniaxial compression along the $x$-axis in an orthorhombic system. In the state optimized for the applied compression with the stress component $T_{xx}$, the spontaneous stress components $T_{yy}$ and $T_{zz}$ will reach zero by expanding the unit cell along $y$ and $z$ axes. The red box represents the initial crystal structure, while the blue box indicates its shape under $x$-axis compression.}
\label{fig:poisson}
\end{figure}
\begin{table*}
\captionof{table}{The stress tensor $\bm{\mathit{T}}$, strain tensor $\epsilon$, and the corresponding conventional unit cell form in the first-principles calculations. The italic-bold $\bm{\mathit{0}}$ in the stress tensor indicate the spontaneous stress components that reach zero at the unit cell optimized for the applied strain denoted in the leftmost column.
$|\xi|$ indicates percent stress. 
Notice that the applied stress is proportional to the strain component $\bm{\mathit{t}}\sim \xi$. The $\nu_{u}$ and $\nu_{b}$ are the uniaxial and biaxial Poisson's ratios, while $\nu_{si}$ and $\nu_{so}$ are coefficients related to the in-plane and out-plane distortion respectively under the shear stress. The $\nu_{u_y}$ and $\nu_{u_z}$ are Poisson's ratios in $y$ and $z$ directions. }
\label{tab:strain}
\begin{tabular}{P{3cm}P{3.5cm}P{4.5cm}P{4.5cm}} 
\hline
\hline
  & Stress tensor &  Strain tensor &  Unit cell form \\
\hline
$\begin{array}{c} \mathrm{Cubic} \\
\mathrm{Uniaxial-}z   \\  
\end{array} $ &   
$ \left( \begin{array}{ccc}  \bm{\mathit{0}}  & 0 & 0  \\ 0 & \bm{\mathit{0}}  & 0 \\ 0 & 0 & \bm{\mathit{t}}  \\ \end{array}\right)$ & 
$ \left( \begin{array}{ccc}  -\nu_{u}\xi  & 0 & 0  \\ 0 & -\nu_{u}\xi  & 0 \\ 0 & 0 & \xi  \\ \end{array}\right)$ & 
$ \left( \begin{array}{ccc} 1-\nu_{u}\xi  & 0 & 0  \\ 0 & 1-\nu_{u}\xi  & 0 \\ 0 & 0 & 1+\xi  \end{array}\right)$\\
$\begin{array}{c} \mathrm{Cubic} \\
\mathrm{Biaxial-}xy   \\  
\end{array} $ &   
$ \left( \begin{array}{ccc}  \bm{\mathit{t}}  & 0 & 0  \\ 0 & \bm{\mathit{t}}  & 0 \\ 0 & 0 & \bm{\mathit{0}}  \\ \end{array}\right)$ & 
$ \left( \begin{array}{ccc}  \xi  & 0 & 0  \\ 0 & \xi  & 0 \\ 0 & 0 & -\nu_{b}\xi  \\ \end{array}\right)$ & 
$ \left( \begin{array}{ccc} 1+\xi  & 0 & 0  \\ 0 & 1+\xi  & 0 \\ 0 & 0 & 1-\nu_{b}\xi  \end{array}\right)$\\
$\begin{array}{c} \mathrm{Cubic} \\
\mathrm{Shear-}xy   \\  
\end{array} $ &   
$ \left( \begin{array}{ccc}  \bm{\mathit{0}}  & \bm{\mathit{t}} & 0  \\ \bm{\mathit{t}} & \bm{\mathit{0}}  & 0 \\ 0 & 0 & \bm{\mathit{0}} \\ \end{array}\right)$ & 
$ \left( \begin{array}{ccc}  \nu_{si}\xi  & \xi & 0  \\ \xi & \nu_{si}\xi  & 0 \\ 0 & 0 & -\nu_{so}\xi  \\ \end{array}\right)$ & 
$ \left( \begin{array}{ccc} 1+ \nu_{si}\xi  & \xi & 0  \\ \xi & 1+ \nu_{si}\xi  & 0 \\ 0 & 0 & 1-\nu_{so}\xi  \end{array}\right)$\\
$\begin{array}{c} \mathrm{Orthorhombic} \\
\mathrm{Uniaxial-}x   \\  
\end{array} $ &   
$ \left( \begin{array}{ccc}  \bm{\mathit{t}} & 0 & 0  \\ 0 & \bm{\mathit{0}}  & 0 \\ 0 & 0 & \bm{\mathit{0}}  \\ \end{array}\right)$ & 
$ \left( \begin{array}{ccc}  \xi  & 0 & 0  \\ 0 & -\nu_{u_y}\xi  & 0 \\ 0 & 0 & -\nu_{u_z}\xi  \\ \end{array}\right)$ & 
$ \left( \begin{array}{ccc} 1+\xi  & 0 & 0  \\ 0 & 1-\nu_{u_y}\xi  & 0 \\ 0 & 0 & 1-\nu_{u_z}\xi  \end{array}\right)$\\
\hline
\hline
\end{tabular}  
\end{table*}
The stress tensor is estimated from the derivative of the total energy $E$ with respect to the partial strain $\epsilon_{ij}$, as described in \cite{Nielsen1985}:
\begin{equation}
\label{eq:stress}
T_{ij} = \frac{1}{V} \frac{\partial E} {\partial \epsilon_{ij}}
\end{equation}
where strain $\epsilon_{ij}$ transforms $r_{i}$ to $r_{i}+\sum_{j}\epsilon_{ij}r_{j}$ and $V$ represents the volume of the unit cell. 
Details regarding the decomposition of the stress tensor and its implementation in first-principles calculation codes are discussed in \cite{Torrent2007, Sharma2018,Kresse1999}. 
Table \ref{tab:strain} presents the stress and strain tensors, along with the conventional unit cell form, for the cubic system under uniaxial-$z$, biaxial-$xy$, and shear-$xy$, as well as the orthorhombic system under uniaxial-$x$ stress. 

A strain applied to a crystal acts on the applied components of stress, but also the spontaneous stress components, located in italic-bold $\bm{\mathit{0}}$ in Table~\ref{tab:strain} and causes a secondary distortion in the crystal to release the spontaneous stress.
Poisson's ratios are defined as the negative ratio of the transverse strain to an applied axial strain to measure the deformation in material in the direction perpendicular to the direction of the applied strain.
For example, in an orthorhombic crystal compressed along the $x$-direction, the Poisson's ratio $\nu_{u_y}$ quantifies how much the material will contract in the $y$-direction, and $\nu_{u_z}$ quantifies the contraction in the $z$-direction. 
More specifically, when subjected to uniaxial compression along the $x$-axis corresponding to the crystal axis $a$, the lattice constants of the orthorhombic crystal reduced $a$ to $a(1+\xi)$ ($\xi<0$ for compression) under stress $T_{xx}$ while simultaneously increasing $b$ to $b(1-v_{\nu_y}\xi)$ and $c$ to $c(1-\nu_{u_z}\xi)$ due to the spontaneous stress components $T_{yy}$ and $T_{zz}$ as illustrated in Fig.~\ref{fig:poisson}.

When performing first-principles calculations of the piezomagnetic effect, it is essential to establish the appropriate unit cell for the applied strain. 
In the calculation, we use the unit cell deformed according to the right-most column of Table~\ref{tab:strain}.
Then, the Poisson's ratio is determined by optimizing the unit cell so that the spontaneous stress tensor becomes zero. This method ensures an accurate investigation of the piezomagnetic effect.

For convenience of comparison, we provide the conversion factors for commonly used units of piezomagnetic coefficients in the literature:
\sisetup{input-digits = 0123456789\pi}
 \begin{align}
& \qty{1}{\muB\per \kilo B}=\qty[per-mode=symbol]{18.9655}{\gauss\per\mega\pascal},\nonumber \\
&  \qty{1}{\per\oersted }=\qty[per-mode=symbol]{4 \pi e7}{\gauss\per\mega\pascal}, \nonumber \\
&  \qty{1}{\per\tesla }=\qty{ e-4}{\per\oersted}=\qty[per-mode=symbol]{4 \pi e3}{\gauss\per\mega\pascal}. \label{eq:unit}
 \end{align}
In the second and third conversions, the units of the inverse magnetic field, $\unit{\per\oersted}$ or $\unit{\per\tesla}$, are used. 
This is because, in experiments, the change in length (or strain) of a material can be measured as a function of the applied magnetic field. 
%
\subsection{Computational details}
We perform first-principles calculations using the Vienna \textit{Ab initio} Simulation Package (VASP)~\cite{vasp} with pseudopotentials based on the projector augmented-wave (PAW) method~\cite{Kresse1999},
and the GGA-PBE exchange-correlation functional~\cite{GGA-PBE}. 
We choose an energy cut-off of 600 eV for the plane-wave basis set and a uniform $k$-point grid of 24 $\times$ 24 $\times$ 24 for Mn$_3A$N and 16 $\times$ 16 $\times$ 16 for Mn$_3X$. 
The Methfessel–Paxton scheme is used to determine the partial occupancy of orbitals with a smearing width of $0.05$ eV. 

Most manganese nitrides Mn$_3A$N ($A$= Ni, Cu, Zn, Ga, Rh, Pd, Ag, In, Sn, Pt, Au, Hg) have the cubic antiperovskite crystal structure~\cite{1978ex,2013ex,2014ex} which belongs to the space group $Pm\bar{3}m$ ($O^1_h$, No.~221).
As discussed later, the two stable AFM orders shown in Fig.~\ref{fig:mn3ancrystal} are nearly degenerate in energy and exhibit similar optimized lattice parameters—3.830\AA, 3.846\AA, 3.863\AA, and 3.858\AA—obtained by optimizing the cubic unit cells for Mn$_3$NiN, Mn$_3$CuN, Mn$_3$ZnN, and Mn$_3$GaN, respectively. These values reproduce the experimental observations~\cite{2014ex} with discrepancies of less than 1.5\%.

Mn$_3$Sn and Mn$_3$Ge crystallize into a hexagonal structure that belongs to the space group $P6_3/mmc$ ($D^4_{6h}$, No.~194)~\cite{cmp2017}.
The difference in lattice constants between the two AFM states in Fig.~\ref{fig:mn3xcrystal} is less than 0.05\%, i.e., the fully relaxed lattice constants of Mn$_3$Sn are $a = 5.567\,(5.568)\,\mathrm{\AA}$ and $c = 4.432\,(4.430)\,\mathrm{\AA}$, while those of Mn$_3$Ge are $a = 5.244\,(5.243)\,\mathrm{\AA}$ and $c = 4.246\,(4.245)\,\mathrm{\AA}$ for AFM1 (AFM2). 
These values differ by about 2\% from the experimentally observed ones~\cite{exp2014m3sn,exp2016m3ge}.
The Mn atoms occupy the $6h$ ($x, 2x, 1/4$) positions, with the optimized values of $x$ being 0.842 for Mn$_3$Sn and 0.835 for Mn$_3$Ge, while the Sn and Ge atoms are located at the $2c$ ($1/3, 2/3, 1/4$) Wyckoff position. 

To discuss the effect of SOC for piezomagnetism, calculations were performed with and without considering the spin-orbit interaction.
The unit-cell lattice parameters for both the unstrained and strained configurations were optimized with SOC, following the procedure described in Sec.~\ref{sec:optimize}. 
The same optimized parameters were then used for SOC-free calculations, leading to changes in spontaneous stress components less than 1\%.
We evaluated magnetization from spin moments since the contribution of orbital moments is negligible for the piezomagnetic effects as discussed in the Appendix~\ref{seq:appendix}.
The piezomagnetic tensor components are estimated from Eq.~\eqref{eqpiezomagi} with the values of magnetization and stress tensor components of the systems obtained from first-principles calculations for the applied strain.
%
\section{Results} \label {sec:result}
\label{sec:results}
\subsection{Anisotropic piezomagnetism in M\lowercase{n}$_3A$N} \label{sec:man}
\begin{figure} 
\centering 
\includegraphics[width=7cm]{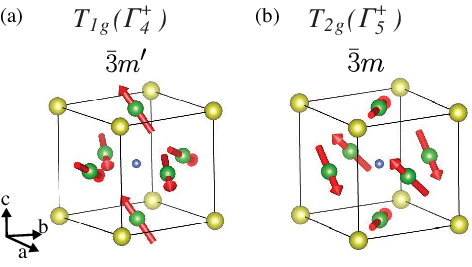}\\
\captionof{figure}{Antiferromagnetic configurations (a) AFM-$T_{1g}$ and (b) AFM-$T_{2g}$ on the cubic Mn$_3A$N crystal. The green, yellow, and blue balls indicate Mn, $A$, and N atoms, respectively. Arrows on Mn atoms indicate the magnetic moments. Magnetic configurations are visualized using VESTA~\cite{Momma:db5098}.}
\label{fig:mn3ancrystal}
\end{figure}
Manganese nitrides, Mn$_3A$N generally stabilize in two distinct non-collinear antiferromagnetic structures, represented by the irreducible representations $T_{1g}(\Gamma^+_4)$ and $T_{2g}(\Gamma^+_5)$~\cite{1978ex,2013ex}. 
We named the two above states as AFM--$T_{1g}$ and AFM--$T_{2g}$ and illustrated them in Fig.~\ref{fig:mn3ancrystal}(a) and (b).
These AFM orders are transformed into each other by spin rotation. 
As a result, the two AFM states are degenerate in the absence of spin–orbit coupling (SOC), and the magnetic anisotropy induced by SOC determines which of the two magnetic orders becomes more stable. 
Indeed, first-principles calculations show the stability of these two magnetic structures with a very small energy difference due to the weak SOC in $3d$ transition metal compounds~\cite{ahe}. 
AFM--$T_{1g}$ and AFM--$T_{2g}$ belong to the magnetic point groups $\bar{3}m'$ and $\bar{3}m$, respectively, with their symmetry operations listed in Table~\ref{tab:symele}.
Consequently, AFM--$T_{1g}$ can induce the anomalous Hall effect, while AFM--$T_{2g}$ prohibits its emergence~\cite{ahe}.
\begin{table}
\captionof{table}{Symmetry operations of magnetic point groups (MP) $\bar{3}m'$ and $\bar{3}m$. $C_{n\mu}$, $I$, $m_{\mu}$  indicates the $n$-fold rotation along the $\mu$ axis, spatial inversion, and mirror operation with the mirror plane normal to the $\mu$ axis, respectively. The prime $(')$ indicates the operations combined with time--reversal operation.}
\label{tab:symele}
\begin{tabular}{P{1.0cm}P{6.0cm}}
\hline
\hline
 MP & Elements   \\
 \hline
$\bar{3}m$ & $E$, $C_{3[111]}$, $C^-_{3[111]}$, $C_{2[1\bar{1}0]}$, $C_{2[01\bar{1}]}$, $C_{2[\bar{1}01]}$, $I$, $IC_{3[111]}$, $IC^-_{3[111]}$, $m_{[1\bar{1}0]}$, $m_{[01\bar{1}]}$, $m_{[\bar{1}01]}$     \\
$\bar{3}m'$ & $E$, $C_{3[111]}$, $C^-_{3[111]}$, $C'_{2[1\bar{1}0]}$, $C'_{2[01\bar{1}]}$, $C'_{2[\bar{1}01]}$, $I$, $IC_{3[111]}$, $IC^-_{3[111]}$, $m'_{[1\bar{1}0]}$, $m'_{[01\bar{1}]}$, $m'_{[\bar{1}01]}$     \\
\hline
\hline
\end{tabular}
\end{table}
\subsubsection{SOC free Mn$_{3}A$N systems}\label{sec:sys_without_SOC_Mn3AN}
\begin{figure*}
\centering
\includegraphics[width=16cm]{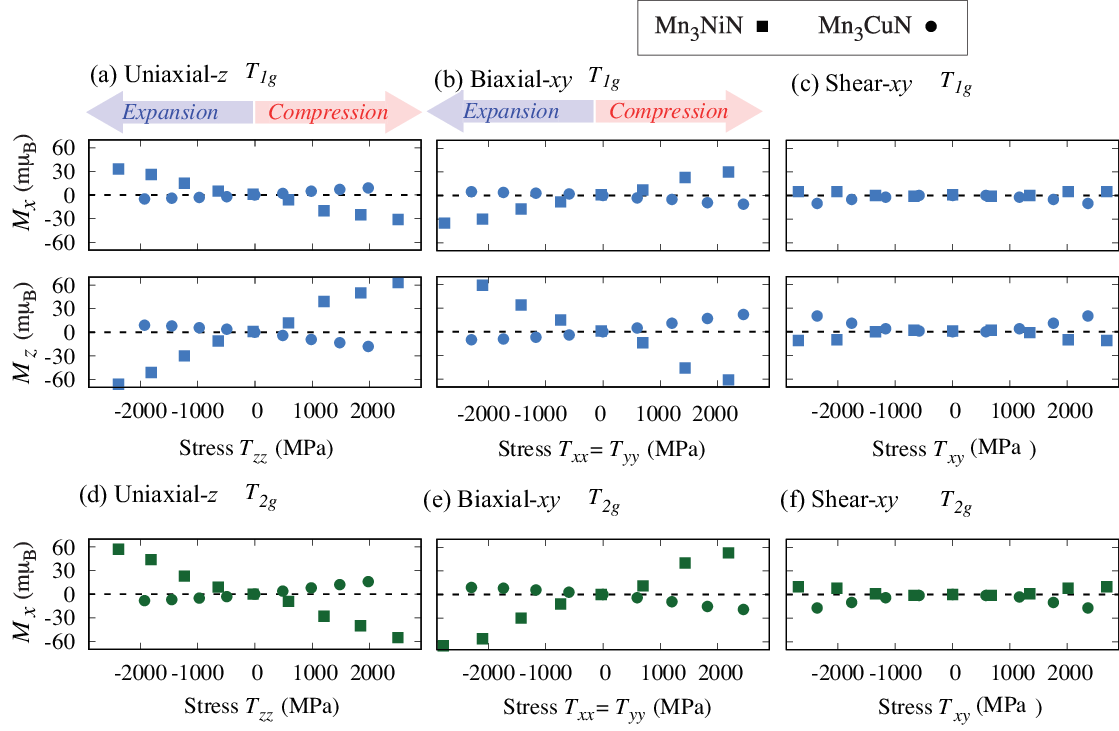}\\
\caption{Magnetization response to uniaxial $z$ compression, biaxial $xy$, and shear $xy$ strain for the AFM-$T_{1g}$ and AFM-$T_{2g}$ states in Mn$_3$NiN and Mn$_3$CuN without SOC.
The terms ``expansion'' and ``compression'' refer to an increase and a decrease in the unit cell volume, respectively.
}
\label{fig:mn3anSOCfree}
\end{figure*}

Here we provide a symmetry argument in the piezomagnetic tensor in the AFM--$T_{1g}$ and AFM--$T_{2g}$ magnetic structures. 
We first consider the SOC--free system, which is invariant under any spin rotation without magnetic orders. 
The symmetries of the magnetically ordered phases in these systems are classified according to their spin point groups and spin space groups~\cite{LITVIN1974538,Litvin:a14103,Liu_PhysRevX.12.021016,Smejkal_PhysRevX.12.031042,Watanabe_PhysRevB.109.094438,Zhenyu_PhysRevX.14.031037,Xiaobing_PhysRevX.14.031038,Yi_PhysRevX.14.031039,Schiff_SciPostPhys.18.3.109}. 
The spin point group $\mathcal{P}$ is generally represented as the direct product of the spin-only point group $\mathcal{P}_{\mathrm{so}}$ and the non-trivial spin point group $\overline{\mathcal{P}}$ and in the present AFM--cases, $\mathcal{P}_{\mathrm{so}}=m$ and  $\overline{\mathcal{P}}={}^{2_x}4/{}^1m\,{}^{3_z}\bar{3}\,{}^{2_x}2/{}^{2_x}m$ (No.~587 in Litvin's Table)~\cite{Litvin:a14103}.  
The spin point group symmetry might impose additional constraints on the physical response tensors to those under the magnetic point group~\cite{Watanabe_PhysRevB.109.094438}. 
The numbers of independent components of the piezomanetic tensor $Q_{ijk}$ under the magnetic point groups $\bar{3}m$ and $\bar{3}m'$ are two and four~\cite{itcD157}, respectively, as shown later.
\setlength{\extrarowheight}{.5ex}
\begin{table}
\captionof{table}{Relationship between magnetization and stresses that takes into account up to linear piezomagnetic effects for AFM-$T_{1g}$ and AFM-$T_{2g}$ states without SOC. From Eq.~\eqref{eq:relation}, $Q_{12}=-\sqrt{3}Q'_{12}$.}
\label{tab:magres_woSOC}
\begin{tabular}{ p{2cm}  p{3.5cm}}
\hline
\hline 
\multicolumn{2}{c}{AFM-$T_{1g}$} \\
\hline 
Uniaxial $z$  &  $M_x = M_y =  Q'_{12} T_{zz}$ \\ 
        & $M_z = -2Q'_{12}T_{zz}  $ \\
Shear $xy$ &  $ M_x = M_y = M_z=0 $\\
Biaxial $xy$ & $M_x = M_y = -Q'_{12} T_{xx}$ \\  
&$M_z = 2Q'_{12}T_{xx} $ \\
\hline
\multicolumn{2}{c}{AFM-$T_{2g}$} \\
\hline 
Uniaxial $z$  &  $M_x = -M_y =  -Q_{12} T_{zz}$ \\ 
        & $M_z = 0  $ \\
Shear $xy$ &  $ M_x = M_y = M_z=0 $\\
Biaxial $xy$ & $M_x = -M_y = Q_{12} T_{xx}$ \\ 
&$M_z = 0 $ \\
\hline
\hline
\end{tabular}  
\end{table}
In the absence of the SOC, however, the symmetry characterized by the combination of the $\pi$ spin  rotation along $[111]$--axis and the time-reversal, offers the following relation: 
\begin{align}
    Q_{xjk}+Q_{yjk}+Q_{zjk}=0. \label{eq:spg_Q_Mn3AN_1}
\end{align}
The mirror operations on $(100)$, $(010)$, and $(001)$--planes in real space provide another relations: 
\begin{align}
Q_{iyz}=Q_{izx}=Q_{ixy}=0.\label{eq:spg_Q_Mn3AN_2}
\end{align}
Equations~\eqref{eq:spg_Q_Mn3AN_1} and \eqref{eq:spg_Q_Mn3AN_2} pose the relations in several components of the piezomagnetic tensor in the AFM--$T_{1g}$ and $T_{2g}$ states in addition to those in the magnetic point groups considering SOC discussed in Sec.~\ref{sec:sys_with_SOC_Mn3AN}.
As a result, the piezomagnetic tensors in each AFM-state are given as follows:
\begin{align}
 \mathbf{Q}^{T_{1g}}&=
\begin{bmatrix}
-2Q'_{12} & Q'_{12} & Q'_{12} & 0 & 0 & 0  \\
Q'_{12} & -2Q'_{12} & Q'_{12} & 0 & 0 & 0 \\
Q'_{12} & Q'_{12} & -2Q'_{12} & 0 & 0 & 0   
\end{bmatrix}
, \label{eq:Q_Mn3AN_T1g}\\
\mathbf{Q}^{T_{2g}}&=
\begin{bmatrix}
0 & Q_{12} & -Q_{12} & 0 &0 & 0  \\
-Q_{12} & 0 & Q_{12} & 0 & 0 &0 \\
Q_{12} & -Q_{12} & 0 & 0 & 0 & 0   
\end{bmatrix}, 
\label{eq:Q_Mn3AN_T2g}
 \end{align}
where we add prime to the components of the piezomagnetic tensor in the AFM--$T_{1g}$ state for the distinction from those in the AFM--$T_{2g}$ state. 
The linear piezomagnetic coefficients for shear stresses $T_{yz}$, $T_{zx}$, and $T_{xy}$ vanish both for the $T_{1g}$- and $T_{2g}$-AFM cases without the SOC.  
The linear development of magnetization for the uniaxial $z$, biaxial $xy$, and shear $xy$ stresses is summarized in Table~\ref{tab:magres_woSOC}.
We note that \emph{nonlinear} piezomagnetic effects can emerge even though the linear piezomagnetic coefficients are prohibited in the system without SOC as such nonlinear piezomagnetic effects have recently been discussed in collinear altermagnets~\cite{Ogawa_2024arXiv241219158O}. 
In the present noncollinear AFM cases, the nonlinear piezomagnetism can appear under shear stresses since the following phenomenological free energy is allowed in the AFM--$T_{1g}$ state: 
\begin{align}
    F&\propto (-M_x-M_y+2M_z)(-T^2_{yz}-T^2_{zx}+2T^2_{xy})  \nonumber \\
    &+ 3(M_x-M_y)(T^2_{yz}-T^2_{zx}), \label{eq:fene_T1g_woSOC}
\end{align}
and in the AFM--$T_{2g}$ state, 
\begin{align}
    F&\propto -(M_x-M_y)(-T^2_{yz}-T^2_{zx}+2T^2_{xy})  \nonumber \\
    &+ (-M_x-M_y+2M_z)(T^2_{yz}-T^2_{zx}) . \label{eq:fene_T2g_woSOC}
\end{align}
When considering SOC, additional terms appear in Eqs.~\eqref{eq:fene_T1g_woSOC} and \eqref{eq:fene_T2g_woSOC}. 

Figure~\ref{fig:mn3anSOCfree} shows the stress dependence of magnetization for $T_{1g}$- and $T_{2g}$-AFM states.
We can observe several characteristic behaviors described by the piezomagnetic tensors in Eqs.~\eqref{eq:Q_Mn3AN_T1g} and \eqref{eq:Q_Mn3AN_T2g}. 
The magnetization develops linearly for the uniaxial $z$ and biaixal $xy$ stresses. 
In the AFM--$T_{1g}$ state, the relation $M_z\simeq -2M_x$ holds for the uniaxial $z$ stress $T_{zz}$ consistent with Eq.~\eqref{eq:Q_Mn3AN_T1g}.
In the AFM--$T_{2g}$ state, the relation $M_x \simeq M_y$  holds for the uniaxial $z$ stress $T_{zz}$; therefore, the plot with regard to $M_y$ is not presented here. 
Figures~\ref{fig:mn3anSOCfree}(c), (f) and Fig.~\ref{fig:mn3cunSOC} for smaller range of magnetization, show magnetization as a function of shear stress $T_{xy}$.  
In contrast to the cases of the uniaxial $z$ and biaixal $xy$ stresses, the magnetization behaves quadratically for $T_{xy}$ and the relation $\bm{M}(+T_{xy})=\bm{M}(-T_{xy})$ holds as discussed above. 
This is a stark contrast to the finite SOC cases, in which the piezomagnetic coefficients $\bm{Q}$ for the shear stress remain finite, as discussed in Sec.~\ref{sec:sys_with_SOC_Mn3AN}.  

Now, let us discuss the relation between the piezomagnetic tensors in the AFM-$T_{1g}$ and $T_{2g}$ states. 
In the SOC free cases, the Hamiltonian in the AFM-$\Gamma$ ($\Gamma=T_{1g}$, $T_{2g}$) states may be written as follows: 
\begin{align}
\mathcal{H}_{\Gamma} = \mathcal{H}_{0} +  \mathcal{H}_{\mathrm{MF},\Gamma}, 
\end{align}
 where $\mathcal{H}_{0}$ is the Hamiltonian in the paramagnetic state and 
$\mathcal{H}_{\mathrm{MF},\Gamma}$ represents the molecular-field of the AFM--$\Gamma$ states, respectively. 
We note that $\mathcal{H}_{0}$ is invariant under any spin rotation and its combination with the time-reversal symmetry $\theta$ due to the absence of the SOC. 
On the other hand, the molecular field Hamiltonian of the AFM--$T_{1g}$ state, $\mathcal{H}_{\mathrm{MF},T_{1g}}$, and that of the AFM--$T_{2g}$ state, $\mathcal{H}_{T_{2g}}$, can be related with each other by the spin--rotation such as the four--fold rotation of spin along the $[111]$--direction $C^{(\mathrm{sp})}_{4[111]}$ as follows:
\begin{align}
\mathcal{H}_{\mathrm{MF},T_{1g}} = C^{(\mathrm{sp})}_{4[111]}\mathcal{H}_{\mathrm{MF},T_{2g}}\left(C^{(\mathrm{sp})}_{4[111]}\right)^{-1}.
\end{align}
Linear piezomagnetic coefficients $Q_{ijk}$ are transformed as $Q'_{ijk}=\sum_{i'}D_{ii'}(C^{(\mathrm{sp})}_{4[111]})Q_{i'jk}$.
We thus obtain the following relation between piezomagnetic tensors in the AFM--$T_{1g}$ and $T_{2g}$ states: 
\begin{align}
Q_{12}=-\sqrt{3}Q'_{12}. 
\label{eq:relation}
\end{align}
From the results in Table~\ref{tab:magres_woSOC} and Eq.~\eqref{eq:relation}, 
the net magnetization in the AFM-$T_{1g}$ and -$T_{2g}$ states develops with the same magnitude, but in a different direction, rotated $\pm\pi/2$ along the [111] axis.
Therefore, the magnetization in the AFM--$T_{1g}$ and $T_{2g}$ states up to the linear order of $T_{zz}$ are given as follows: 
\begin{align}
\bm{M}^{T_{1g}}
&=-\frac{1}{\sqrt{3}}(1,1,-2)Q_{12}T_{zz}, \label{eq:M_T1g_Tzz_woSOC} \\
\bm{M}^{T_{2g}}
&= (-1,1,0)Q_{12}T_{zz}. \label{eq:M_T2g_Tzz_woSOC}
\end{align}
Indeed, our numerical results for the SOC-free cases, for example, $M_z^{T_{1g}} = 0.064\mu_{\mathrm{B}}$ and $M_x^{T_{2g}} = 0.055\mu_{\mathrm{B}}$ at uniaxial-$z$ compression $T_{zz}$= 2500 $\mathrm{GPa}$ for Mn$_3$NiN holds the relation $M_z^{T_{1g}} \simeq -\frac{2}{\sqrt{3}}M_x^{T_{2g}}$ from Eqs.~\eqref{eq:M_T1g_Tzz_woSOC} and \eqref{eq:M_T2g_Tzz_woSOC}. 
This relation holds in weak SOC systems, as discussed in Sec.~\ref{sec:sys_with_SOC_Mn3AN}. 
\subsubsection{Mn$_{3}A$N systems with SOC}\label{sec:sys_with_SOC_Mn3AN}
\begin{figure*} 
\centering 
\includegraphics[width=16cm]{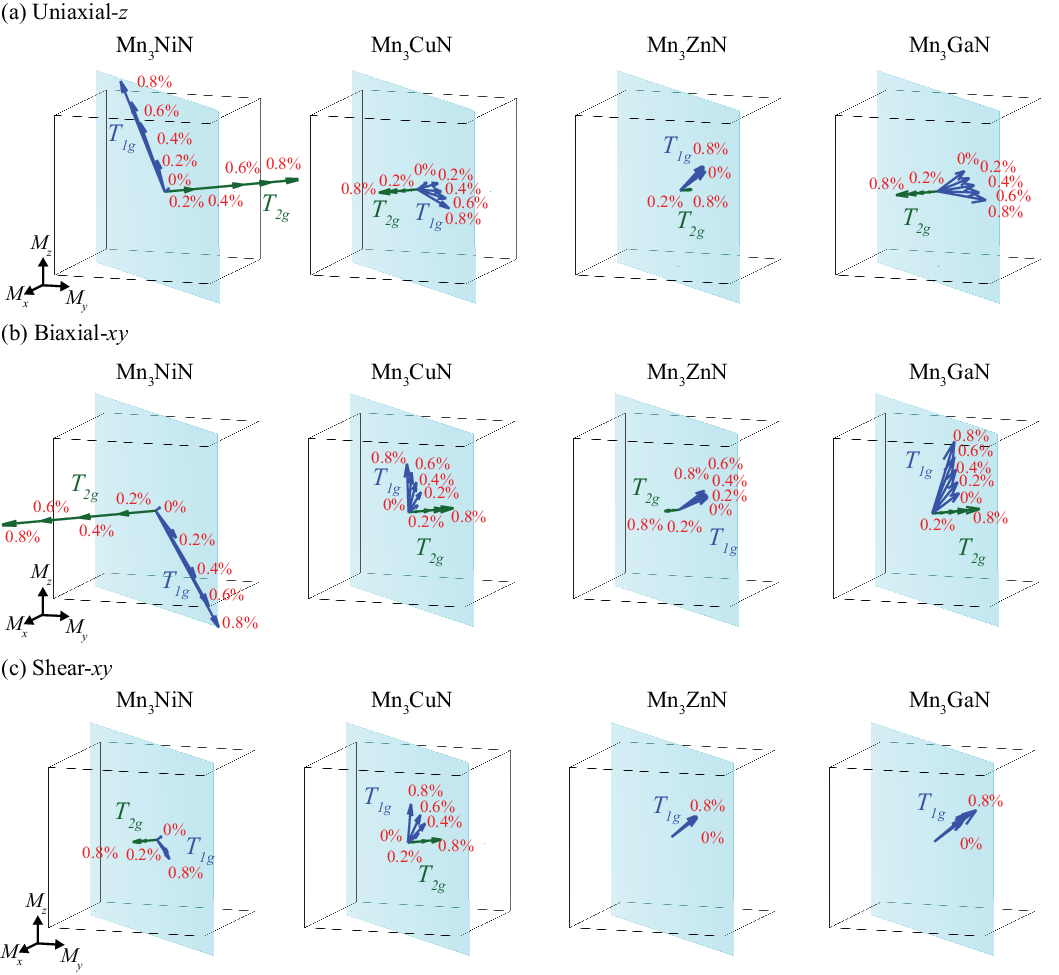}\\
\captionof{figure}{Anisotropic piezomagnetic response for the AFM--$T_{1g}$ state (dark-blue arrows) and AFM--$T_{2g}$ state (dark-green arrows) and AFM--$T_{1g}$ state (dark-blue arrows) in Mn$_3A$N with SOC. The light-blue planes indicate $(\bar{1}10)$ planes, with red numbers indicating the percent stress. In the AFM--$T_{2g}$ state, magnetization develops along the $[1\bar{1}0]$ direction, following the relation $M_x=-M_y$ and $M_z=0$ (see Table~\ref{tab:magres}), while in the AFM--$T_{1g}$ state, the magnetization lies in the $(\bar{1}10)$ plane due to the relation $M_x=M_y\neq M_z$ (see Table~\ref{tab:magres}). 
 }
\label{fig:direction}
\end{figure*}
\begin{figure*} 
\centering 
\includegraphics[width=16cm]{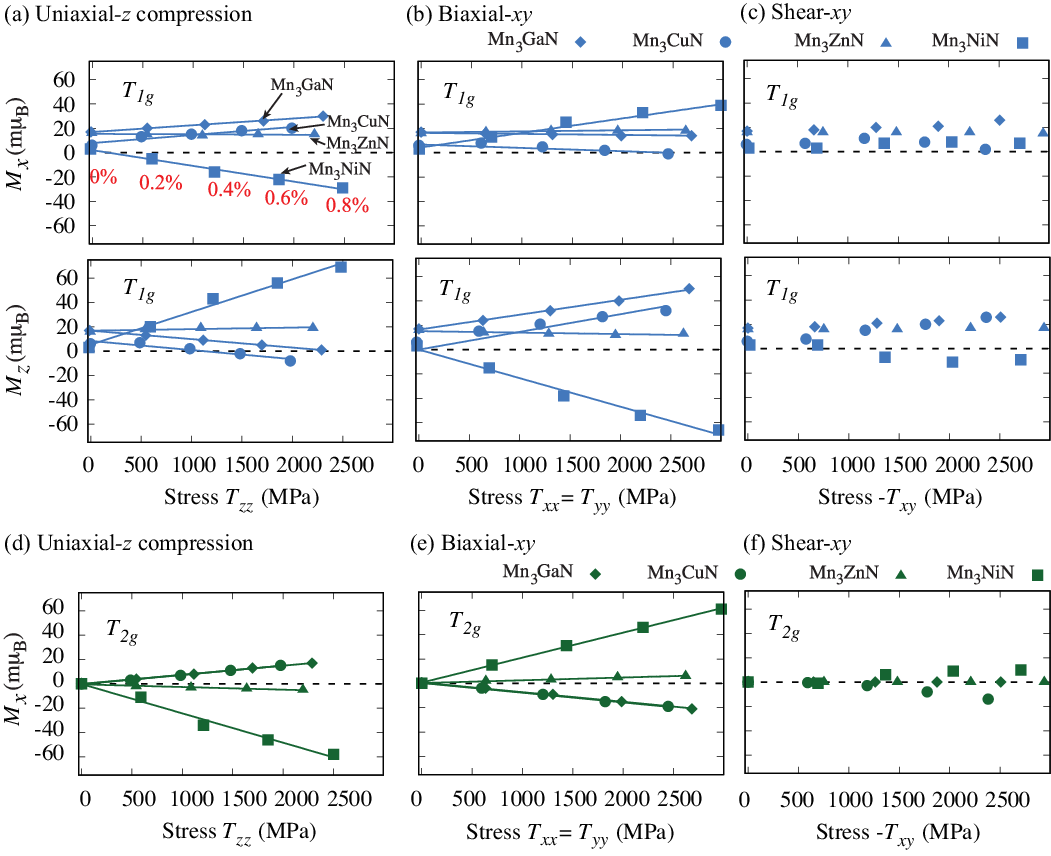}\\
\captionof{figure}{(a)--(f) Magnetization response to the uniaxial $z$ compression, biaxial $xy$, and shear $xy$ for AFM-$T_{1g}$ and AFM-$T_{2g}$ states in Mn$_3A$N with SOC. The calculated data points are shown as markers, and lines represent linear fitting, with red numbers indicating the percent stress. 
No linear fits were applied to the shear stress data, which exhibits a dominant quadratic dependence of the magnetization on stress.}
\label{fig:mn3angather}
\end{figure*}
\begin{figure}
\centering
\includegraphics[width=8cm]{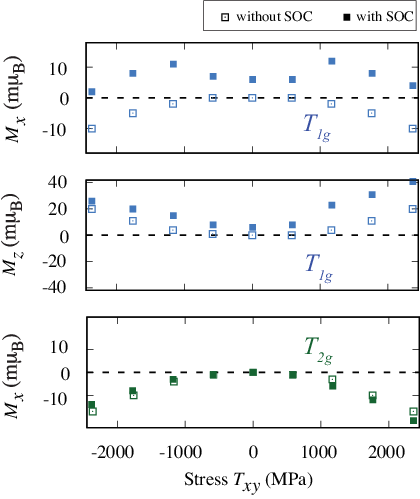}\\
\caption{Magnetization response to shear-$xy$ stress for AFM-$T_{1g}$ and AFM-$T_{2g}$ states with and without SOC in Mn$_3$CuN in a negative and positive range of stress.}
\label{fig:mn3cunSOC}
\end{figure}
\begin{figure*} 
\includegraphics[width=16.0cm]{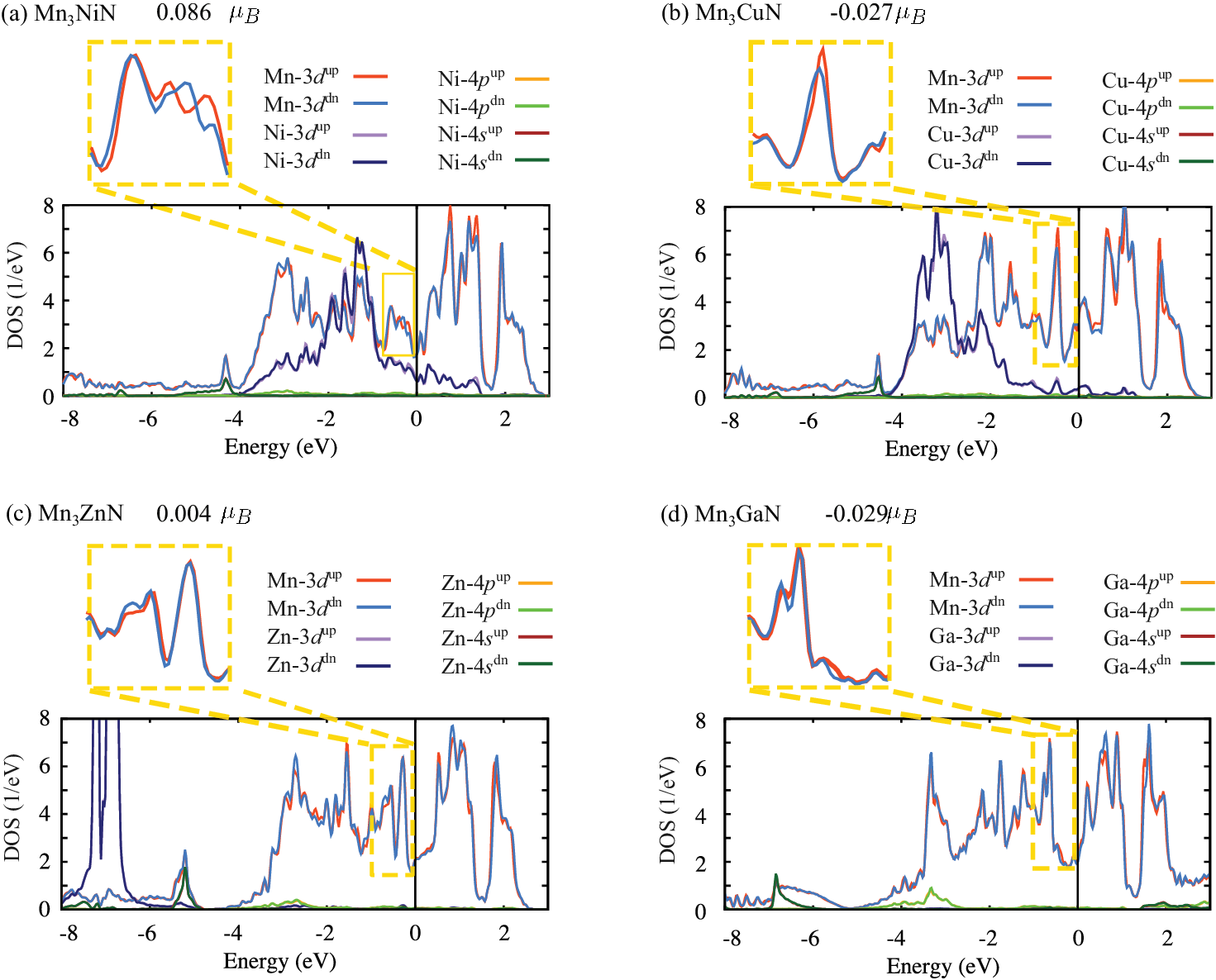}\\
\captionof{figure}{Spin projected density of states at $0.8\%$ biaxial $xy$ stress of AFM-$T_{2g}$ state in series Mn$_3A$N with SOC. Spin up and down are defined along the $[1\bar{1}0]$ quantization axis, which corresponds to the direction of induced magnetization under the $xy$-biaxial strain.}
\label{fig:dos}
\end{figure*}
When the SOC is included, the piezomagnetic tensors for the magnetic point groups $\bar{3}m'$ and $\bar{3}m$ are as follows:
\begin{equation}
\label{3mp}
\mathbf{Q}^{T_{1g}}_{\bar{3}m'}=\left[ \begin{matrix}
Q'_{11} & Q'_{12} & Q'_{12} & Q'_{14} & Q'_{15} & Q'_{15}  \\
Q'_{12} & Q'_{11} & Q'_{12} & Q'_{15} & Q'_{14} & Q'_{15} \\
Q'_{12} & Q'_{12} & Q'_{11} & Q'_{15} & Q'_{15} & Q'_{14}   \\
\end{matrix}
 \right] 
\end{equation}
and
\begin{equation}
\label{3m}
\mathbf{Q}^{T_{2g}}_{\bar{3}m} 
=\left[ \begin{matrix}
0 & Q_{12} & -Q_{12} & 0 & Q_{15} & -Q_{15}  \\
-Q_{12} & 0 & Q_{12} & -Q_{15} & 0 & Q_{15} \\
Q_{12} & -Q_{12} & 0 & Q_{15} & -Q_{15} & 0   \\
\end{matrix}
 \right]
\end{equation}
in the cubic axes setting.
The development of magnetization up to linear order in uniaxial $z$, biaxial $xy$, and shear $xy$ stresses for AFM-$T_{1g}$ and -$T_{2g}$ states with SOC is summarized in Table \ref{tab:magres}. 
The AFM--$T_{1g}$ ($T_{2g}$) states belong to the magnetic point group $\bar{3}m'$ ($\bar{3}m$) and reduce it to $2'/m'$ ($2/m$) for the AFM--$T_{1g}$ ($T_{2g}$) states under uniaxial, biaxial, and shear stresses, as listed in Table~\ref{tab:reducedmn3an}.
Consequently, the magnetization for the unstrained AFM--$T_{2g}$ magnetic structure is zero, while it is finite for the AFM--$T_{1g}$ one. 
The magnetization of the AFM--$T_{1g}$ structure under strain is given by $\bm{\mathit{M = m_0 + QT}}$, where $\bm{\mathit{m_0}}$ represents the magnetization of the unstrained AFM--$T_{1g}$ state.
\begin{table}
\captionof{table}{Relationship between magnetization and
stresses that takes into account up to linear piezomagnetic effects for AFM--$T_{1g}$ and AFM-$T_{2g}$ states with SOC. The $m_{0}$ represents the spontaneous magnetization component for the unstrained AFM--$T_{1g}$ state.}
\label{tab:magres}
 For AFM-$T_{1g}$ \\
\begin{tabular}{P{2cm}P{5cm}}
\hline
\hline
Uniaxial $z$  & $ \left\{\begin{array}{l} M_x = M_y = m_{0}+ Q'_{12} T_{zz} \\  M_z =m_{0}+Q'_{11}T_{zz} \end{array}\right. $ \\
Shear $xy$ &  $\left\{\begin{array}{l}  M_x = M_y = m_{0} + Q'_{15} T_{xy}   \\  M_z = m_{0}+Q'_{14} T_{xy}  \end{array}\right. $\\
Biaxial $xy$ & $\left\{\begin{array}{l} M_x = M_y = m_{0}+ (Q'_{11}+ Q'_{12})T_{xx} \\  M_z = m_{0}+Q'_{12}(T_{xx}+ T_{yy})  \end{array}\right. $ \\
\hline
\hline
\end{tabular}  \\
 For AFM-$T_{2g}$ \\
\begin{tabular}{P{2.5cm}P{4.5cm}}
\hline
\hline
Uniaxial $z$ & $\left\{ \begin{array}{l}  M_x = -M_y = - Q_{12} T_{zz}   \\  M_z = 0 \end{array}\right. $ \\
Shear $xy$ & $\left\{\begin{array}{l}  M_x = -M_y =Q_{15} T_{xy}  \\  M_z = 0 \end{array}\right. $ \\
Biaxial $xy$ & $\left\{ \begin{array}{l}  M_x = -M_y = Q_{12} T_{xx}   \\  M_z = 0 \end{array}\right. $ \\
\hline
\hline
\end{tabular}
\end{table}
\setlength{\extrarowheight}{.5ex}
\begin{table}
\captionof{table}{Magnetic point groups of AFM--$T_{1g}$ and AFM--$T_{2g}$ magnetic structures and of those under applied strains in Mn$_3A$N. The symbols with prime indicate the operations combined with time-reversal.}
\label{tab:reducedmn3an}
\begin{tabular}{P{4.0cm}P{1.5cm}P{1.5cm}}
\hline
\hline
&  \multicolumn{2}{c} {Magnetic symmetry} \\
Type of strain & AFM-$T_{1g}$ & AFM-$T_{2g}$ \\ 
\hline
No strain & $\bar{3}m'$  & $\bar{3}m$  \\
Uniaxial & $2'/m'$ & $2/m$ \\
Biaxial & $2'/m'$ & $2/m$\\
Shear & $2'/m'$ & $2/m$\\
Triaxial & $\bar{3}m'$ & $\bar{3}m$\\
\hline
\hline
\end{tabular}
\end{table}
\begin{table}
\captionof{table}{Calculated magnetization under strain in Mn$_3$NiN and  Mn$_3$GaN comparing with other theoretical (Thr.) and experiment (Exp.) works.}
\label{tab:com}
\begin{tabular}{P{3.5cm}C{1.5cm}C{2.8cm}} 
\hline
\hline
& \multicolumn{2}{c} {Magnetization $(\mu_B)$}  \\
 & This work & References  \\
 \hline
$\circ$  Mn$_3$NiN  \\ 
1$\%$ uniaxial (AFM--$T_{2g}$) & $-0.099$  & $-0.1$ ~\cite{pmtheory201796} $^{*}$Thr. \\
0.2$\%$ biaxial (AFM--$T_{1g}$) & 0.024 & $\numrange[range-phrase=\text{--}]{0.05}{0.12}$ ~\cite{exp2018pmmn3nin} Exp.\\
$\circ$  Mn$_3$GaN \\
1$\%$ biaxial (AFM--$T_{2g}$) & 0.037 & $0.040$~\cite{pmtheory2008} Thr. \\
\hline
\hline
\multicolumn{3}{l}{$^{*}$\footnotesize{The value with fixed Poisson's ratio in Ref.~\cite{pmtheory201796}.}} \\
\end{tabular}  
\end{table}
\begin{table}[h!]
\centering
\captionof{table}{Computed piezomagnetic coefficients in the unit $\unit[per-mode=symbol]{\gauss\per\mega\pascal}$.}
\label{tab:pm}
 AFM-$T_{1g}$ magnetic structures ($\bar{3}m'$)\\
\setlength{\tabcolsep}{10pt}
\begin{tabular}{l r r r r}
\hline
 & Ni & Cu & Zn & Ga \\
\hline
\( Q'_{11} \) & \(0.051\) & \(-0.014\) & \(-0.002\) & \(-0.013\) \\
\( Q'_{12} \) & \(-0.025\) & \(0.013\) & \(-0.001\) & \(0.010\) \\
\hline
\end{tabular}
 \\
 ~
 \\
 AFM-$T_{2g}$ magnetic structures ($\bar{3}m$)\\
\setlength{\tabcolsep}{10pt}
\begin{tabular}{l r r r r}
\hline
 & Ni & Cu & Zn & Ga \\
\hline
\( Q_{12} \) & \(0.044\) & \(-0.015\) & \(0.004\) & \(-0.014\)\(^*\) \\
\hline
\end{tabular}
$^{*}$\footnotesize{The value in \cite{pmtheory2008} for Mn$_3$GaN is $\qty[per-mode=symbol]{0.038}{\gauss\per\mega\pascal}$}.
\end{table}

We present the calculated magnetization under uniaxial $z$, shear $xy$, and biaxial $xy$ strains in Fig.~\ref{fig:direction}. The magnetization develops for the stresses according to Table~\ref{tab:magres}.
For AFM-$T_{2g}$ states, the magnitude of magnetization increases with the direction kept as the applied stress increases.
Whereas for AFM-$T_{1g}$ states, the magnetization lies and rotates within the $(\bar{1}10)$ plane under strain. 
The rotation of the magnetization for AFM-$T_{1g}$ states, clearly seen in Mn$_3$CuN and Mn$_3$GaN in Fig.~\ref{fig:direction}, is attributed to SOC, since 
the magnetization in the SOC-free case is parallel to $[11\bar{2}]$-direction for any type of stresses except shear stress in the present study according to Eq.~\eqref{eq:Q_Mn3AN_T1g} [for example, see Eq.~\eqref{eq:M_T1g_Tzz_woSOC} and Table~\ref{tab:magres_woSOC} for the $T_{zz}$ case].  
On the other hand, magnetization in Mn$_3$NiN for AFM-$T_{1g}$ state with relatively weak SOC is nearly pointed to $[\bar{1}\bar{1}2]$ ($[11\bar{2}]$)--direction under the uniaxial $z$ (biaxial $xy$) stress as for the SOC-free case shown in Eq.~\eqref{eq:M_T1g_Tzz_woSOC} . 
This implies that the origin of the piezomagnetism in Mn$_3$NiN is not the SOC but the exchange interaction as discussed in Ref.~\cite{pmtheory201796}. 
In contrast to collinear AFMs, in which the  SOC is essential for piezomagnetism~\cite{MORIYA195973,Spaldin_PhysRevX.14.011019}, the exchange interaction-driven piezomagnetic effects are characteristic in noncollinear magnets.

Table~\ref{tab:com} lists the magnetization under stress obtained in this work and from previous theoretical and experimental reports.
Our calculated magnetizations are comparable to previous theoretical and experimental reports, as seen in Table~\ref{tab:com}.
Magnetization components under various strains are shown in Figs.~\ref{fig:mn3angather}(a)--(f),
therein we used the linear fitting to plot solid lines to see the linear development of magnetization under stress. 
We also present the magnetization response to strain under both positive and negative shear stress for Mn$_3$CuN (see Fig.~\ref{fig:mn3cunSOC}). 
The magnitude and direction of magnetization in response to stress show a similar dependence on the applied strain as observed in the case without SOC described in Sec.~\ref{sec:sys_without_SOC_Mn3AN}, but with small deviations originating from SOC. For example, there is a slight deviation from the relation $\bm{M}(+T_{xy}) = \bm{M}(-T_{xy})$ as shown in Fig.~\ref{fig:mn3cunSOC}. 
Magnetization clearly exhibits a quadratic behavior for $T_{xy}$, as discussed above.

The results indicate that the magnitude and direction of the induced magnetization in Mn$_3A$N strongly depend on the type of applied stress and the underlying magnetic structure, revealing an anisotropic piezomagnetic effect.
This anisotropy enables efficient control of magnetization in Mn$_3A$N through mechanical stress and tuning of the magnetic structure via chemical composition or external magnetic fields, offering a novel strategy for magnetization manipulation.

Moreover, the anisotropic piezomagnetic response under strain can serve as a sensitive probe for identifying the magnetic phase.
Magnetic structures are typically characterized using neutron scattering and/or NMR measurements.
However, in complex situations, such as those involving a low symmetry of the magnetic point group and the presence of off-diagonal magnetoelectric tensor elements, these methods may not fully identify the magnetic structures.
In such cases, cross-correlated phenomena, such as the piezomagnetic effect, can serve as complementary methods to determine the magnetic structures. 
For example, the magnetoelectric effect, a typical cross-correlated response, was used in combination with neutron diffraction in pulsed magnetic fields to determine the magnetic structures of LiFePO$_4$ in its high field phase~\cite{janas}. 

The calculated piezomagnetic tensor components are listed in Table~\ref{tab:pm}. 
The piezomagnetic coefficient $Q_{12}$ for Mn$_3$GaN in the AFM-$T_{2g}$ state calculated in this work, \qty[per-mode=symbol]{-0.014}{\gauss\per\mega\pascal}, is
smaller than that reported in a previous theoretical study, $q=\qty{3e-10}{\per\oersted }=\qty[per-mode=symbol]{0.038}{\gauss\per\mega\pascal}$ in the absolute values~\cite{pmtheory2008}\footnote{Ref.~\cite{pmtheory2008} did not offer a detailed explanation of the piezomagnetic tensor components.}.
Table~\ref{tab:pm} indicates the largest piezomagnetic effect in Mn$_3$NiN within the investigated compounds. Meanwhile, the relationship of Eq.~\eqref{eq:relation} satisfied without SOC is most closely observed in Mn$_3$NiN, indicating the smallest effect of SOC among the compounds.

To understand the contribution of the electron orbital to the induced magnetization under strain, we plot the spin project density of states for AFM--$T_{2g}$ at $0.8\%$ biaxial--$xy$ strain in Fig.~\ref{fig:dos}. 
Spin up and down are defined for the $[1\bar{1}0]$ quantization axis, which is the direction of induced magnetization under the biaxial--$xy$ strain. The spin splitting of the Mn-$3d$ orbital shows a significant effect in Fig.~\ref{fig:dos}, indicating that the magnetic moments of the Mn-$3d$ states play a crucial role in the piezomagnetic effect in Mn$_3A$N.
%
\subsection{Anisotropic piezomagnetism in M\lowercase{n}$_3X$}  \label{sec:mn3x}
\begin{figure} 
\centering 
\includegraphics[width=7cm]{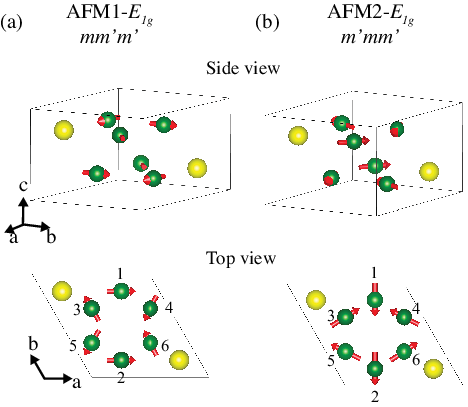}\\
\captionof{figure}{Antiferromagnetic configurations (a) AFM1 and (b) AFM2 on the hexagonal Mn$_3X$ crystal. The green and yellow balls indicate Mn and $X$ atoms, respectively. Arrows on Mn atoms indicate the magnetic moments. Magnetic configurations are visualized using VESTA~\cite{Momma:db5098}.}
\label{fig:mn3xcrystal}
\end{figure}
In Mn$_3X$, applying magnetic fields along the $x$ and $y$--axes is known to stabilize the AFM1 and AFM2 magnetic structures~\cite{tomiyoshi}, respectively, as shown in Figs.~\ref{fig:mn3xcrystal}(a) and \ref{fig:mn3xcrystal}(b).
The AFM1 (AFM2) belongs to the orthorhombic magnetic point group of $mm'm'$ ($m'mm'$), which contains the symmetry operations $E$, $C_{2x}$, $C'_{2z}$, $C'_{2y}$, $I$, $m_x$, $m'_z$, and $m'_y$ ($E$, $C_{2y}$, $C'_{2z}$, $C'_{2x}$, $I$, $m_y$, $m'_z$, and $m'_x$).
In terms of magnetic representation theory, 
the AFM1 and AMF2 magnetic structures form the independent bases of the $E_{1g}$ irreducible representation under $D_{6h}$ crystallographic point group~\cite{cmp2017}.
Similar to the AFM--$T_{1g}$ and AFM--$T_{2g}$ magnetic structures considered in Mn$_3A$N, the AFM1 and AFM2 magnetic structures in Mn$_3X$ are transformed each other by spin rotation and are energetically degenerate in the absence of spin-orbit interaction.
\subsubsection{SOC free Mn$_{3}X$ systems}
\label{sec:wosocMn3X}
\begin{figure}
\centering
\includegraphics[width=7cm]{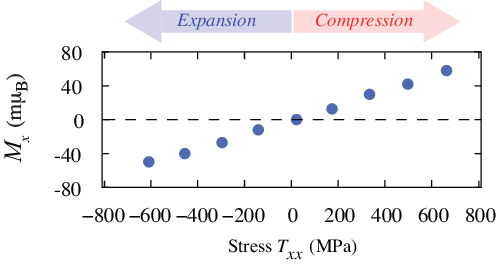}\\
\caption{Magnetization response to uniaxial $x$ strain for the AFM1 state in Mn$_3$Sn without SOC.}
\label{fig:mn3xnosoc}
\end{figure}
In the same manner as in the case of Mn$_3$$A$N, we provide piezomagnetic tensors in the SOC free system.  
The AFM1 and AFM2 states are conjugate subgroups of the parent group for the paramagnetic state and both are described by the same spin point group $\mathcal{P}=\mathcal{P}_{\mathrm{so}}\times \overline{\mathcal{P}}$ with $\mathcal{P}_{\mathrm{so}}=2/m$ and $\overline{\mathcal{P}}={}^{3_z}6/{}^1m\,{}^{2_x}m\,{}^{2_{xy}}m$ (No.~498 in Litvin's table)~\cite{Litvin:a14103}.
The higher symmetry in the system without the SOC than in the magnetic point groups $mm'm'$ and $m'mm'$ reduces the independent components of several physical response tensors. 

The piezomagnetic tensors in the AFM1 and AFM2 states in the absence of the SOC are respectively given as follows:
\begin{align}
&    \mathbf{Q}^{\mathrm{AFM1}}
= \begin{bmatrix}
Q_{11} & -Q_{11} & 0 & 0 & 0 & 0  \\
0 & 0 & 0 & 0 & 0 & 2Q_{11}   \\
0 & 0 & 0 & 0 & 0 & 0     \\
\end{bmatrix}, \label{eq:Q_AFM1_Mn3X_woSOC}\\
&    \mathbf{Q}^{\mathrm{AFM2}}
= \begin{bmatrix}
0 & 0 & 0 & 0 & 0 & 2Q'_{22}   \\
-Q'_{22} & Q'_{22} & 0 & 0 & 0 & 0  \\
0 & 0 & 0 & 0  & 0& 0     \\
\end{bmatrix}. \label{eq:Q_AFM2_Mn3X_woSOC}
    \end{align}
The $\pm \pi/2$ spin rotation along the $[001]$-direction, which transforms the AFM1 and AFM2 states into each other, leads to the following relation in the piezomagnetic tensor components: 
\begin{align}
Q_{11}=Q'_{22}.
\label{eq:relation2}
\end{align}
Figure~\ref{fig:mn3xnosoc} shows the $T_{xx}$-dependence of magnetization $M_x$ for the AFM1 state in Mn$_3$Sn.
$M_x$ develops linearly against $T_{xx}$ and vanishes at $T_{xx}=0$. 
Zero net magnetization without strain is a consequence of the present spin point group symmetry of the SOC-free case, in contrast to finite magnetization of the system with the SOC as discussed next, in which the net magnetization can emerge.
%
\subsubsection{Mn$_{3}X$ systems with SOC}
\label{sec:wsocMn3X}
\begin{figure} 
\centering 
\includegraphics[width=7cm]{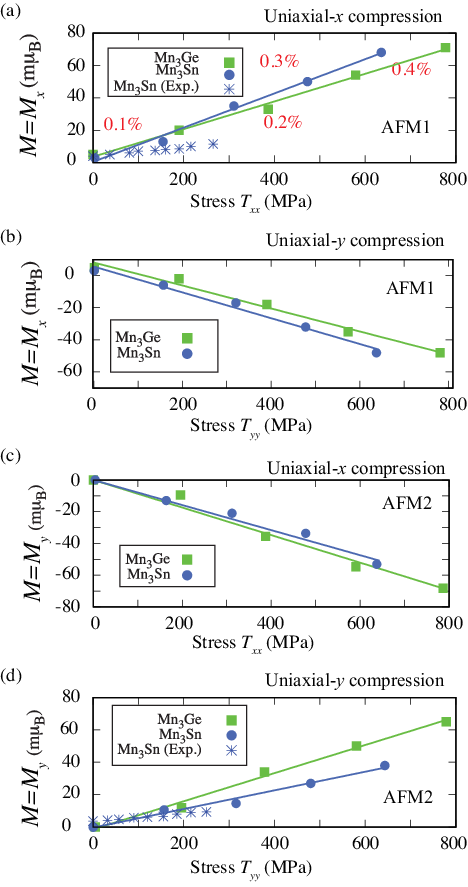}\\
\captionof{figure}{(a)--(c) Magnetization response to the compression in Mn$_3$Sn and Mn$_3$Ge with SOC: calculated data points are shown as markers, and lines represent linear fitting, with red numbers indicating the percent stress. The blue stars show experiment values taken from \cite{exp2022pmahc} for Mn$_3$Sn.}
\label{fig:mn3xgather}
\end{figure}
 \begin{figure} 
\centering 
\includegraphics[width=8cm]{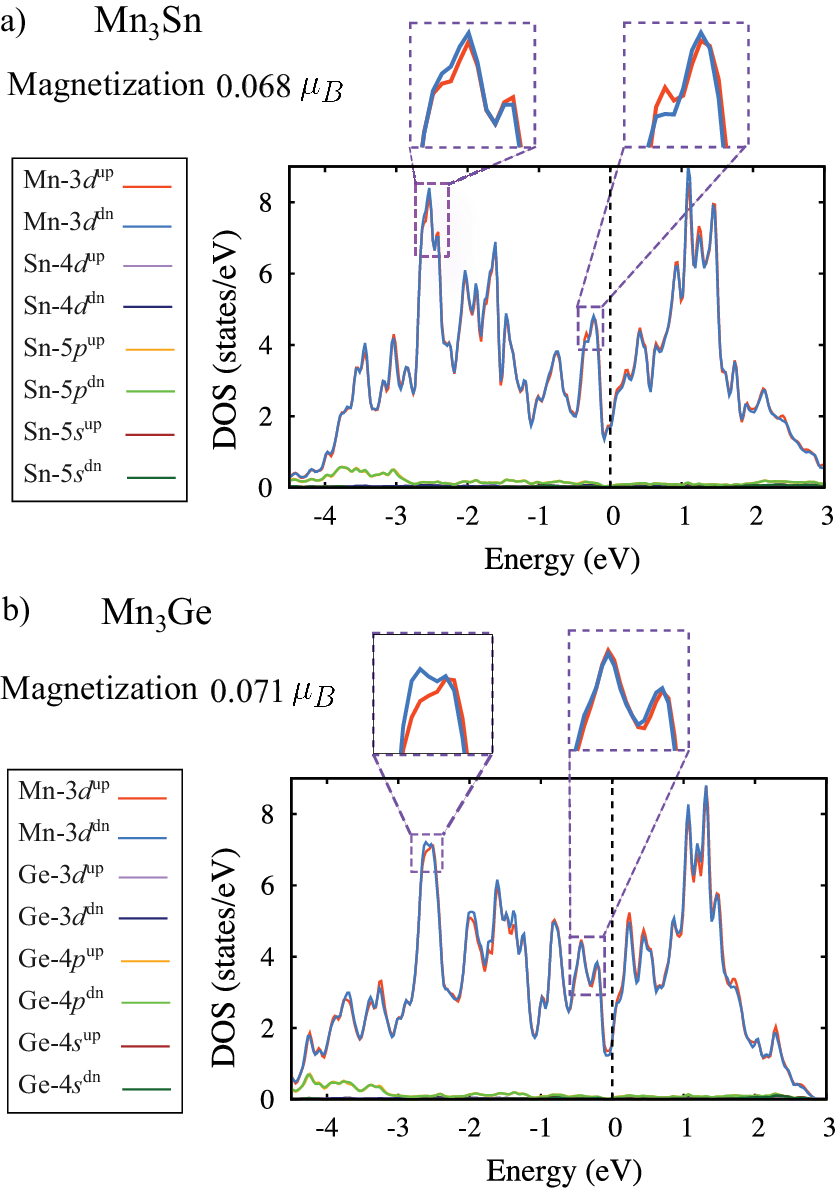}\\
\captionof{figure}{Spin projected density of states at $0.4\%$ uniaxial-$x$ compression for AFM1 state of Mn$_3$Sn (a) and Mn$_3$Ge (b) with SOC.
Spin up and down are defined for the $x$-quantization axis.}
\label{fig:mn3xdos}
\end{figure}
\begin{figure} 
\centering 
\includegraphics[width=8cm]{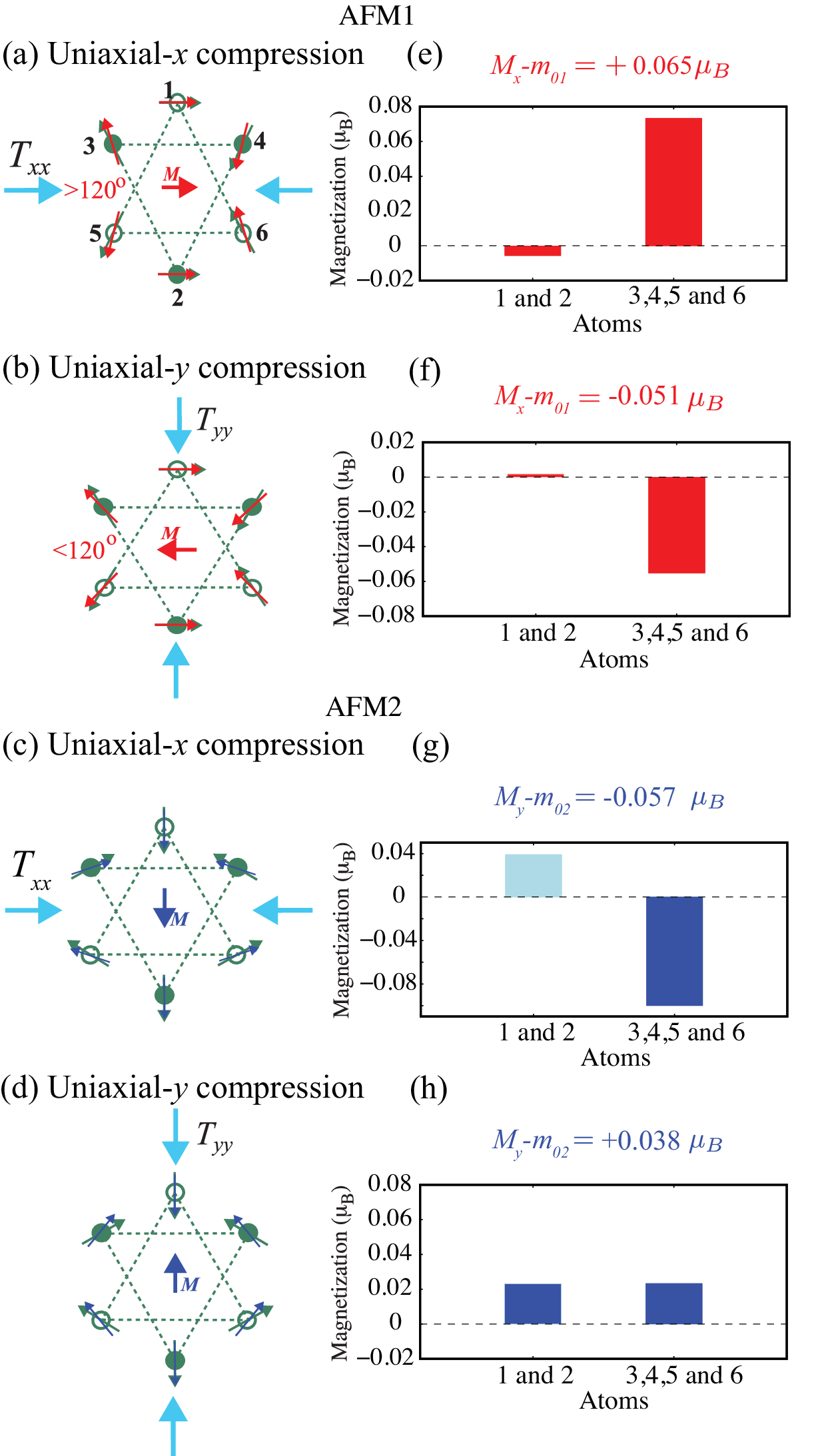}\\
\captionof{figure}{(a)--(d) Illustration of changes in magnetic moments under uni-axial compression in the $x$ and $y$ directions for AFM1 and the $y$ directions for AFM2 in Mn$_3$Sn with SOC. The green arrows indicate magnetic moments at unstrained states and the red (blue) arrows show magnetic moments under compression for AFM1 (AFM2) states. (e)--(h) Contribution of Mn-$d$ states to magnetization on atoms 1 \& 2 and on atoms 3, 4, 5 \& 6 under $0.4 \%$ uniaxial-$x$ and\% uniaxial-$y$ compression for AFM1 and AFM2. The formulas above each figure indicate the magnetization under $x$-compression and $y$--compression relative to the unstrained state.}
\label{fig:mn3xorigin}
\end{figure}
The piezomagnetic tensors of Mn$_3X$ are in the following form for AFM1: 
\begin{equation}
\label{eq:afm1tensor}
\mathbf{Q}^{\mathrm{AFM1}}_{{mm'm'}}
=\left [ \begin{matrix}
Q_{11} & Q_{12} & Q_{13} & 0 & 0 & 0  \\
0 & 0 & 0 & 0 & 0 & Q_{26}   \\
0 & 0 & 0 & 0 & Q_{35} & 0     \\
\end{matrix}
 \right ]
\end{equation}
and for AFM2:
\begin{equation}
\label{eq:afm2tensor}
\mathbf{Q}^{\mathrm{AFM2}}_{{m'mm'}}
=\left [ \begin{matrix}
0 & 0 & 0 & 0 & 0 & Q'_{16}   \\
Q'_{21} & Q'_{22} & Q'_{23}& 0 & 0 & 0  \\
0 & 0 & 0 & Q'_{34}  & 0& 0     \\
\end{matrix}
 \right ].
\end{equation}

From the form of the piezomagnetic tensors, we can write the magnetization up to the linear order of stress components for AFM1: 
\begin{equation}
\label{eq:afm1}
\left \{ \begin{matrix}
M_x &=& m_{01} + Q_{11} T_{xx} + Q_{12} T_{yy} + Q_{13} T_{zz}  \\
M_y &=& Q_{26} T_{xy}   \\
M_z &=& Q_{35} T_{xz}     \\
\end{matrix}
\right. ,
\end{equation}
and for AFM2:
\begin{equation}
\label{eq:afm2}
\left \{ \begin{matrix}
M_x &=& Q'_{16} T_{xy}   \\
M_y &=& m_{02} + Q'_{21} T_{xx} + Q'_{22} T_{yy} + Q'_{23} T_{zz}  \\
M_z &=& Q'_{34} T_{yz}     \\
\end{matrix}
\right .
,
\end{equation}
where $m_{01}$ and $m_{02}$ are magnetization for the unstrained states of AFM1 and AFM2, respectively.
For Mn$_3$Sn, our calculations yield a value of $m_{01}=\qty{0.003}{\mu_{\mathrm{B}}}$ along the same direction as atoms 1 and 2 in Fig.~\ref{fig:mn3xcrystal}, which is in agreement with the experimental data~\cite{exp2022pmahc}. 
Since net magnetization is prohibited without SOC for the unstrained AFM states as discussed in Sec.~\ref{sec:wosocMn3X}, the emergence of finite magnetization is a consequence of the finite SOC.
Equations (\ref{eq:afm1}) and (\ref{eq:afm2}) describe the anisotropic piezomagnetic effect in Mn$_3X$, showing the development of magnetization depending on the magnetic structures and the type of applied stress.

Since the magnetic orders of AFM1 and AFM2 in Mn$_{3}X$ already break the hexagonal crystal symmetry into orthorhombic symmetry, the applied uniaxial strain along the orthorhombic axes $x$, $y$, and $z$ does not further break the magnetic point group symmetry. 
Therefore, AFM1 develops magnetization only in the $x$-direction under uniaxial strain $T_{xx}$ as $M_x =m_{01}+ Q_{11} T_{xx}$, under $T_{yy}$ as $M_x = m_{01}+Q_{12} T_{yy}$, and under $T_{zz}$ as $M_x =m_{01}+ Q_{13} T_{zz}$ (see Eq.~\ref{eq:afm1}). 
Similarly, AFM2 develops magnetization solely in the $y$-direction under the uniaxial strain $T_{xx}$, $T_{yy}$, or $T_{zz}$ as $M_y =m_{02}+ Q_{21} T_{xx}$, $M_y = m_{02}+Q_{22} T_{yy}$, and $M_y =m_{02}+ Q_{23} T_{zz}$, respectively.
Table~\ref{tab:reducedmn3x} presents a supplementary list that illustrates the magnetic point group under strain. 
\begin{table}
\captionof{table}{Magnetic point groups of AFM1 (AFM2) magnetic order and of those under applied strains in Mn$_{3}X$.}
\label{tab:reducedmn3x}
\begin{tabular}{P{4cm}P{4cm}}
\hline
\hline
Type of strain &  Magnetic symmetry  \\
 \hline
No strain & $mm'm'$ ($m'mm'$) \\
Uniaxial $x,y,z$ & $mm'm'$ ($m'mm'$) \\
Biaxial $xy,yz,zx$ & $mm'm'$ ($m'mm'$) \\
Shear $xz$ or $yz$ & $\bar{1}$ ($\bar{1}$)\\
Shear  $xy$  & $2'/m'$ ($2'/m'$)\\
\hline
\hline
\end{tabular}
\end{table}
\begin{table}
\centering
\captionof{table}{Computed piezomagnetic coefficients in the unit $\unit[per-mode=symbol]{\gauss\per\mega\pascal}$.}
\label{tab:pmmn3x}
\setlength{\tabcolsep}{14pt}
\begin{tabular}{l r r r }
\hline
\hline
 &  &   Mn$_3$Sn & Mn$_3$Ge \\
\hline
AFM1 &\( Q_{11} \) & \(0.200\) &\(0.162\) \\
 &\( Q_{12} \)  & \(-0.152\) & \(-0.136\) \\
AFM2 &\( Q'_{21} \)  & \(-0.150\) & \(-0.165\) \\
      &\( Q'_{22} \)  & \(0.109\) & \(0.165\) \\
\hline
\hline
\end{tabular}
$^{*}$\footnotesize{The experiment value for Mn$_3$Sn is $\qty[per-mode=symbol]{0.055}{\gauss\per\mega\pascal}$~\cite{exp2022pmahc}}.
\end{table}

In our work, we focus on the piezomagnetism of the uniaxial stress for Mn$_3X$, which is experimentally observed~\cite{exp2022pmahc}.
The computed magnetizations developed under strain for AFM1 and AFM2 are shown in Figs.~\ref{fig:mn3xgather}(a)--(c) along with the experimental data in Ref.~\cite{exp2022pmahc}. 
Linear fitting lines are used for estimating piezomagnetic coefficients.
The larger computed magnetization compared to experiments can be understood by the temperature difference, \textit{i.e.}, our first-principles calculations based on the density functional theory are for the ground state (zero Kelvin) while the experiment observes the magnetization at room temperature. 
As a result, for the AFM1 state, $Q_{11}=\qty[per-mode=symbol]{0.200}{\gauss\per\mega\pascal}$ calculated for Mn$_3$Sn is higher than the experimental data $\qty[per-mode=symbol]{0.055}{\gauss\per\mega\pascal}$ in Ref.~\cite{exp2022pmahc} (See Table~\ref{tab:pmmn3x}).
The $Q_{11}$ value for the AFM1 state estimated for Mn$_3$Ge from our calculations is $\qty[per-mode=symbol]{0.162}{\gauss\per\mega\pascal}$, which is comparable to that for Mn$_3$Sn. 
The opposite signs of $Q_{11}$ and $Q_{12}$  correspond to the opposite change in magnetization observed in Fig.~\ref{fig:mn3xgather} (a) and (b) where applying compression in the $x$-direction results in positive magnetization and applying compression in the $y$-direction results in negative magnetization. 
The relations $Q_{11}=-Q_{12}$ of Eq.~\eqref{eq:Q_AFM1_Mn3X_woSOC} and $-Q'_{21}=Q'_{22}$ of Eq.~\eqref{eq:Q_AFM2_Mn3X_woSOC} are satisfied in the SOC-free limit and hold approximately even in the finite SOC case due to the weak SOC effects in Mn$_3X$, leading to a sign reversal between $Q_{11}$ and $Q_{12}$ and between $Q'_{21}$ and $Q'_{22}$.
The ability to control signs and directions of magnetization through strain in the case of Mn$_3X$ demonstrates a significant advantage for spintronics applications.
For the AFM2 state, the values $Q'_{22}$ are $\qty[per-mode=symbol]{0.109}{\gauss\per\mega\pascal}$ and $\qty[per-mode=symbol]{0.165}{\gauss\per\mega\pascal}$ for Mn$_3$Sn and Mn$_3$Ge, respectively, indicating a significant piezomagnetic effect in these materials.
The deviation from the relationship $Q_{11}=Q'_{22}$ in Eq.~\eqref{eq:relation2} is attributed to the SOC.

\begin{table}
\captionof{table}{Size of magnetic moments for unstrained, $0.4\%$ compression in $x$ and $y$ directions in unit magneton Bohr ($\mu_B$) for AFM1 and AFM2 states of Mn$_3$Sn.}
\label{tab:mn3xmoments}
(a) For AFM1
\begin{tabular}{P{2.2cm}P{1.6cm}P{1.8cm}P{1.8cm}}
\hline
\hline
Mn atoms &  Unstrain & $0.4\%$ uniaxial-$x$ compression & $0.4\%$ uniaxial-$y$ compression  \\
 \hline
Atoms 1,2 & 2.900 & 2.891 & 2.895 \\
Atoms 3,4,5,6 & 2.899 & 2.894 & 2.891 \\
\hline
\hline
\end{tabular}
(b) For AFM2
\begin{tabular}{P{3.0cm}P{2.2cm}P{2.2cm}}
\hline
\hline
Mn atoms &  Unstrained & $0.4\%$ uniaxial-$y$ compression  \\
 \hline
Atoms 1,2 & 2.896 & 2.889 \\
Atoms 3,4,5,6 & 2.894 & 2.891  \\
\hline
\hline
\end{tabular}
\end{table}

We present the projected density of states for the AFM1 state of Mn$_3X$ ($X$ = Sn and Ge) in Fig.~\ref{fig:mn3xdos}. The spin-up and spin-down states are defined along the $x$-axis, which is the magnetization direction under in-plane stress in the AFM1 state. Similar to Mn$_3A$N, the spin-projected density of states in Mn$_3$Sn and Mn$_3$Ge reveals that the dominant contribution to induced magnetization comes from the splitting of Mn-$3d$ states in Mn$_3X$.

We illustrate the change in magnetic moments under uniaxial compression in the $x$- and $y$-directions for AFM1 and AFM2 in Table~\ref{tab:mn3xmoments} and Figs.~\ref{fig:mn3xorigin}(a)--(d). 
The flip of magnetic moments on all Mn sites with applying uniaxial pressure is reported in Ikhlas \textit{et al}.~\cite{exp2022pmahc}.
In the current calculations, applying the uniaxial stress breaks no magnetic point group symmetry and keeps the time-reversal pair of the AFM states energetically equivalent. 
As a result, the experimentally observed spin flip under uniaxial compression cannot be directly reproduced within our investigation.

Here, we discuss the origin of the magnetization under uniaxial compression for the AFM states in Mn$_{3}X$.
From Table~\ref{tab:mn3xmoments}, we can see that uniaxial compression slightly reduces the magnetic moments at all Mn atoms.
As shown in Figs.~\ref{fig:mn3xorigin}(a)-(d), our calculations indicate that uniaxial compression preserves the direction of the magnetic moments at atoms 1 and 2, 
while it rotates the magnetic moments at atoms 3, 4, 5, and 6.
In the magnetic order of AFM1, the $y$-components of all atoms cancel out. For $x$-compression, as shown in Fig.~\ref{fig:mn3xorigin}(a), 
the magnetic moments of atoms 3, 4, 5, and 6 rotate to reduce partial magnetization along $x$-axis, \textit{i.e.}, the angles between the moments of atoms 3--5 and 4--6 increase to more than $120^\circ$, resulting in positive magnetization in total. 
In contrast, for AFM1 under $y$--compression, the magnetic moments of atoms 3, 4, 5, and 6 rotate to increase the partial magnetization along the $x$-axis, \textit{i.e.}, the angles between the moments of atoms 3--5 and 4--6 decrease to less than $120^\circ$, resulting in negative magnetization in total.

Figures~\ref{fig:mn3xorigin}(e--h) show the decomposition of magnetization into the contribution of Mn atoms for uniaxial $0.4\%$ $x$- and $y$- compression for AFM1 and AFM2. The result shows that the rotation of magnetic moments on atoms 3, 4, 5, and 6 significantly contributes to the magnetization induced under stress.
\section{Conclusion} \label{conclusion}
We investigated the anisotropic responses of magnetization for stress due to piezomagnetic effect of Mn$_3A$N ($A$=Ni, Cu, Zn, and Ga) and Mn$_3X$ ($X$= Sn and Ge), focusing on the stable magnetic structures transformed into each other by spin rotation, \textit{i.e.}, the AFM-$T_{1g}$ and $T_{2g}$ states for Mn$_3A$N and the AFM1 and AFM2 states in Mn$_3X$, respectively. 
The detailed symmetry analysis and first-principles calculations for the AFM states with and without SOC reveal the magnetic exchange and anisotropic contribution of the piezomagnetic effects in these antiferromagnetic compounds.

For Mn$_3A$N, magnetization develops in two distinct directions under the same stress applied to stable AFM states with the differences in magnitude attributed to SOC. 
The stress dependence of the magnetization obtained for Mn$_3$NiN is almost the same as that without the SOC, indicating the weak SOC effect in Mn$_3$NiN. 
Thus, the large piezomagnetic effect in Mn$_3$NiN, explaining the experimental observation~\cite{exp2018pmmn3nin}, is mainly driven by the exchange interaction. 
We also showed that the SOC causes the rotation of net magnetization on the $(\bar{1}10)$ plane only for the AFM-$T_{1g}$ states and found the large magnetization-rotation effects in Mn$_3$CuN and Mn$_3$GaN, which have the SOC relatively larger than Mn$_3$NiN, for the applied stresses.

For Mn$_3X$, the uniaxial stress along the $x$ and $y$ directions breaks no magnetic point group symmetry, and the magnetization for the AFM1 and AFM2 states develops along the $x$ and $y$ directions, respectively, for the uniaxial stresses. Meanwhile, the sign of the induced magnetization depends on the direction of stress, and its dependence is explained from the exchange interaction from the analysis of the piezomagnetic tensor for systems without SOC.

While piezomagnetic effects are typically observed in the collinear AFM phases, those in noncollinear AFM states produce fruitful phenomena such as quadratic development, rotation, and stress type dependence of the induced magnetization. 
We showed that spin point group theory as well as ordinary magnetic point group theory in combination with first-principles calculation is efficient for the analysis of the piezomagnetic effects and other cross-correlated responses.
These findings demonstrate that the signs and directions of magnetization in Mn$_3A$N and Mn$_3X$ can be precisely controlled through strain and magnetic fields, offering potential applications in strain-tunable magnetic devices based on noncollinear antiferromagnets.
%
\section*{Acknowledgement}
This research is supported by JSPS KAKENHI Grants Numbers JP19H01842, JP20H05262, JP20K05299, JP20K21067, JP21H01789, JP21H04437, JP23K20824, JP23K25827, JP23H01130, JP24K00581, JP24K00588, and by JST PRESTO Grant Number JPMJPR17N8. We also acknowledge the use of the supercomputing system, MASAMUNE-IMR, at CCMS, IMR, Tohoku University in Japan.

\appendix
\section{Orbital magnetization} \label{seq:appendix}
%
\begin{figure} 
\includegraphics[width=7cm]{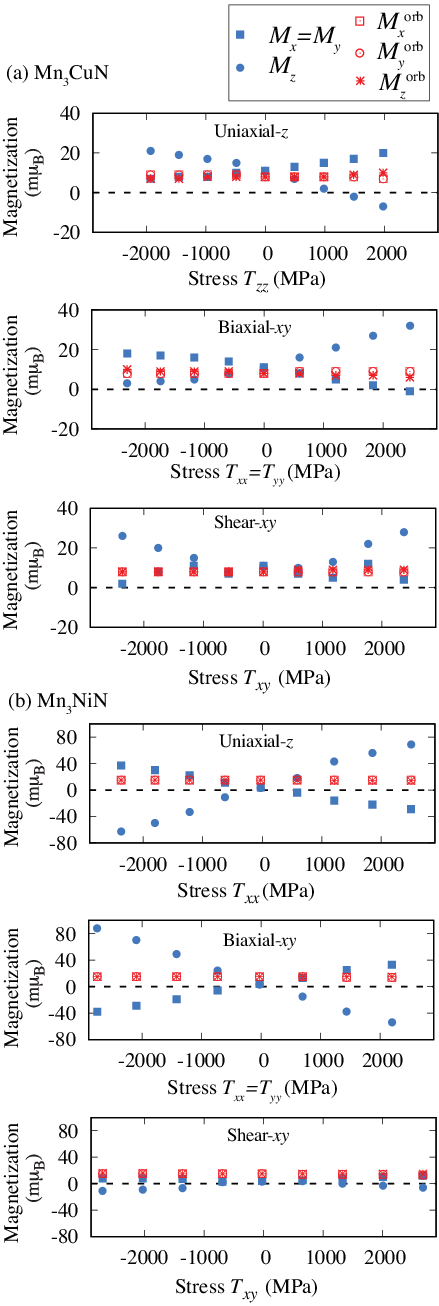}\\
\captionof{figure}{The orbital magnetization $M^{\mathrm{orb}}_x, M^{\mathrm{orb}}_y, M^{\mathrm{orb}}_z$ (red) and spin magnetization $M_x, M_y, M_z$(blue) under uniaxial, biaxial, and shear strain for AFM-$T_{1g}$ state of Mn$_3$CuN and Mn$_3$NiN with SOC.}
\label{fig:mn3cunorbital}
\end{figure}
 \begin{figure} 
\centering 
\includegraphics[width=7cm]{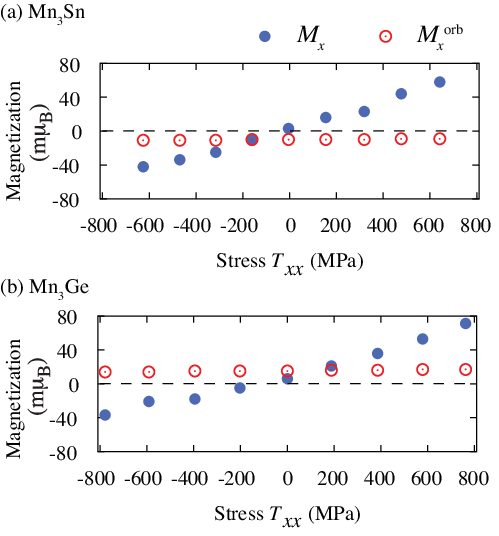}\\
\captionof{figure}{The orbital magnetization $M^{\mathrm{orb}}_x$ (red) and spin magnetization $M_x$ (blue) under uniaxial--$x$ strain for AFM1 states of Mn$_3$Sn (a) and Mn$_3$Ge (b) with SOC.}
\label{fig:mn3xorbital}
\end{figure}
In the main text, we discuss the piezomagnetic effect based on calculations of the spin moments. Here, we address the contribution of the orbital moments to magnetization.
In SOC free systems, the net orbital magnetization can emerge only under noncoplanar magnetic orders and vanishes in the noncollinear coplanar magnetic states in Mn$_3A$N and Mn$_3X$~\cite{Watanabe_PhysRevB.109.094438}.
We thus focus on orbital magnetization in systems including the SOC.

Figure \ref{fig:mn3cunorbital} shows the orbital magnetization responses to strain in Mn$_3$CuN.
In this work, we calculate the orbital magnetization as the sum of local orbital moments on each atom, neglecting the itinerant contribution from modern theory~\cite{Xiao_PhysRevLett.95.137204,Thonhauser_PhysRevLett.95.137205,Vanderbilt_book_2018}, which is beyond the scope of the present study.
In the AFM--$T_{1g}$ states, the orbital magnetization already exists with the unstrained condition when SOC is included and remains nearly unchanged under the strain against the development of the spin magnetization.
In the AFM--$T_{2g}$ states, the orbital magnetization is zero in the unstrained condition from the symmetry and remains zero under strain with the weak dependence for the stress as in the case of AFM--$T_{1g}$.

Figure \ref{fig:mn3xorbital} shows the orbital magnetization $M^{\mathrm{orb}}_x$ under the applied strain in Mn$_3$Sn, the $M^{\mathrm{orb}}_y$ and $M^{\mathrm{orb}}_z$ are forbidden from the symmetry.
The orbital magnetization remains nearly unchanged under strain against the development of the spin magnetization as indicated in Fig.~\ref{fig:mn3xorbital}. 

From the above discussion, we safely neglect the contribution of orbital magnetization for the current discussion on piezomagnetism for both Mn$_3A$N and Mn$_3X$.
\section{Posisson's ratio and the unitcell volume under appied strain}
\begin{figure*} 
\centering 
\includegraphics[width=16cm]{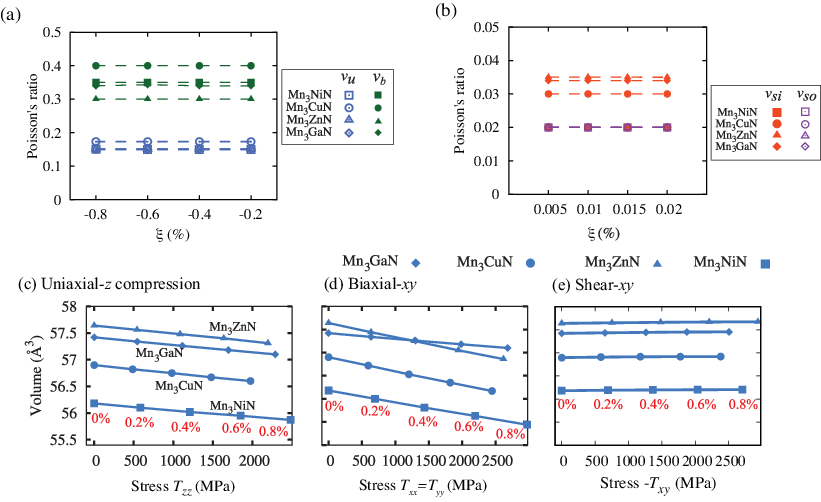}\\
\captionof{figure}{(a) Uniaxial and biaxial Poisson's ratios, $\nu_u$ and $\nu_b$, as a function of percent stress ($\xi$) under uniaxial-$z$ compression and biaxial $xy$ strain. 
(b) Poisson's ratios, $\nu_{si}$ and $\nu_{so}$, corresponding to in-plane and out-of-plane distortions, as a function of percent stress ($\xi$) under $xy$ shear stress.
(c)--(e) Changing in the unit-cell volume under uniaxial-$z$ compression, biaxial $xy$, and shear $xy$ strain in Mn$_3A$N.} 
\label{fig:mn3anpoison}
\end{figure*}
\begin{figure*} 
\centering 
\includegraphics[width=16cm]{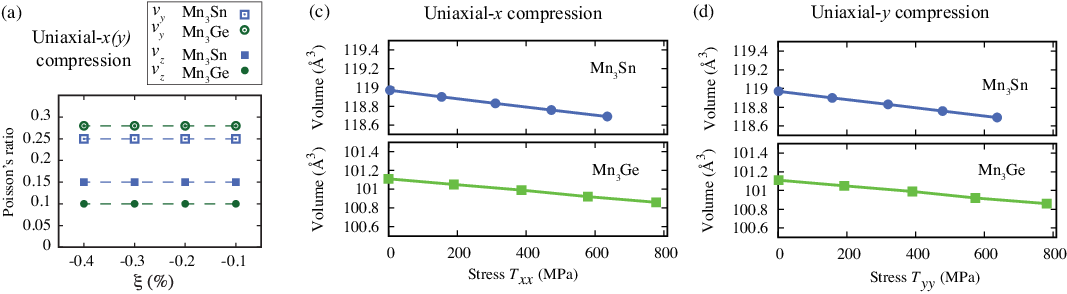}\\
\captionof{figure}{(a) Poisson's ratios $\nu_y$ and $\nu_z$ in the $y$ and $z$ directions under uniaxial-$x$ compression in Mn$_3$Sn and Mn$_3$Ge for AFM1. The parameter $\xi$ represents the percent compression.
(b)--(c) Changing of unit-cell volume under the uniaxial $x(y)$ compression for AFM1.}
\label{fig:mn3xpoisson}
\end{figure*}
Several approximations have been used to study the magnetic response, such as keeping two lattice constants $a,c$ unchanged under the uniaxial stress along $b$ for V$_2$Se$_2$O~\cite{HYMa2021}; keeping volume, or keeping Poisson's ratio for Mn$_3A$N~\cite{pmtheory201796,pmtheory201795,pmtheory2008}. 
However, as shown for cubic InAs in Ref.~\cite{Hammer2007}, the Poisson's ratio can vary depending on the magnitude and orientation of the applied biaxial stress.
In this study, we use first-principles evaluations of the Poisson's ratio by optimizing the stress tensor rather than relying on experimental data or fitting total energies for a fixed AFM order for a range of lattice parameter $(a, a/c)$ to Birch-Murnaghan equation of state as theoretical works by J.~Zemen~\cite{pmtheory201796,pmtheory201795}.
We evaluated the Poisson's ratios after optimizing the stress tensor to investigate the piezomagnetic effect.
The results in Fig.~\ref{fig:mn3anpoison} (a), (b) and Fig.~\ref{fig:mn3xpoisson} (a) show that the Poisson's ratios remain constant under applied strain for both Mn$_3A$N and Mn$_3X$. 
Furthermore, the Poisson's ratios for uniaxial and biaxial stress are much larger than those for shear stress, indicating that uniaxial and biaxial strain induce significantly greater deformation than shear stress.

While Poisson's ratios do not change under strain, the unit-cell volume changes significantly due to the applied strain for both Mn$_3A$N (see Figs.~\ref{fig:mn3anpoison}(c) and (d) under uniaxial and biaxial strains and Mn$_3X$ (see Figs.~\ref{fig:mn3xpoisson}(c) and (d)). 
Under shear-$xy$ strain, the unit-cell volume also increases because of expansion in the 
$z$ direction; however, this increase is much smaller than under uniaxial or biaxial strains (see Fig.~\ref{fig:mn3anpoison}), consistent with the Poisson's ratios discussed above. Examining the volume change under stress thus reveals how much the lattice is compressed or expanded at various stress levels. 
%
%
%
\end{document}